\documentclass[twocolumn,showpacs,preprintnumbers,amsmath,amssymb]{revtex4}

\usepackage{graphicx}
\usepackage{dcolumn}
\usepackage{bm}
\usepackage{amssymb}
\newcommand{\Eqa}[1]{&\hspace{-0.5em}#1\hspace{-0.5em}&}

\begin{document}


\title{Collective quantization of axially symmetric \\gravitating $B=2$ skyrmion}

\author{Hisayuki Sato$^{1}$}
 \email{sugar@dec.rikadai.jp}
\author{Nobuyuki Sawado$^{1}$}%
 \email{sawado@ph.noda.tus.ac.jp}
\author{Noriko Shiiki$^{1},^{2}$}%
 \email{norikoshiiki@mail.goo.ne.jp}

\affiliation{%
$^{1}$Department of Physics, Faculty of Science and Technology, 
Tokyo University of Science, Noda, Chiba 278-8510, Japan\\
$^{2}$Department of Management, Atomi University, Niiza, Saitama 
352-8501, Japan
}%

\date{\today}

\begin{abstract}%
In this paper we perform collective quantization of an axially symmetric skyrmion 
with baryon number two.
The rotational and isorotational modes are quantized to obtain the static 
properties of a deuteron and other dibaryonic objects such as masses, charge 
densities, magnetic moments. We discuss how the gravity affects to those observables. 
\end{abstract}%

\pacs{12.39.Dc, 21.10.-k, 04.20.-q}
\maketitle %
\section{Introduction}
The Skyrme model~\cite{sky} is considered as an unified theory of hadrons 
by incorporating baryons as topological solitons of pion fields, called skyrmions. %
The topological charge is identified as the baryon number $B$. %
Performing collective quantization for a $B=1$ skyrmion, one can obtain 
proton and neutron states within 30$\%$ error~\cite{adkins83}. %

Correspondingly multi-skyrmion solutions are expected to represent 
nuclei~\cite{man,kope,verbaa}.
The static properties of a $B=2$ skyrmion such as %
mass spectra, mean charge radius, baryon-number density, magnetic moment, 
quadrupole moment and transition moment %
were studied in detail by Braaten and Carson upon 
collective zero mode quantization~\cite{bra}. %
The results confirmed that the quantized $B=2$ skyrmion can be interpreted as a deuteron. 

The Einstein-Skyrme (ES) model in which the Skyrme fields coupled to gravity was 
first considered by Luckock and Moss~\cite{moss}. %
They obtained spherically symmetric black hole solutions with Skyrme hair. %
It is the first discovered counter example to the no-hair conjecture. %
Axially symmetric regular and black hole skyrmion solutions with $B=2$ were 
constructed in~\cite{np}. Subsequently the model was extended to the $SU(3)$ and 
higher baryon number with discrete symmetries to study gravitating skyrmion 
solutions~\cite{ioa}. 

In the Einstein-Skyrme theory, the Planck mass is related to the pion decay 
constant $f_{\pi}$ and coupling constant $\alpha$ by $M_{pl}=f_{\pi}\sqrt{4\pi /\alpha}$. 
To realize the realistic value of the Planck mass, the coupling constant should be 
extremely small with $\alpha \sim O(10^{-39})$, %
which makes the effects of gravity on the skyrmion negligible. %
We, therefore, consider $\alpha$ as a free parameter and study the strong coupling limit 
to manifest the effects of gravity on skyrmion spectra. 
The property of the skyrmion solution could be drastically changed for large values of 
$\alpha$. 
Certainly studying the effects of gravity on soliton spectra is interesting itself.  
Nevertheless we further attempt to give an interpretation to the solution as a gravitating 
deuteron or dibaryonic object and study their mass spectra and other static observables. 

The first step towards the study of gravitational effects  
on the quantum spectra of skyrmions was taken in Ref.~\cite{np2} 
by performing collective quantization of a $B=1$ gravitating skyrmion. 
It was shown there that the qualitative change in the mass difference, mean charge 
radius and charge densities under the strong gravitational influence confirms the 
attractive feature of gravity while the reduction of the axial coupling 
and transition moments by the strong gravity indicates the gravitational effects as 
a stabilizer of baryons.

In this paper we extend the work~\cite{np2} to the $B=2$ axially symmetric 
gravitating skyrmion and estimate the static properties of a deuteron or other dibaryonic 
objects. We observe how strong gravity affects the baryon observables {\it e.g.}, 
mass spectra, mean charge radius, baryon-number density, magnetic moment, 
quadrupole moment and transition moment.

Although the Skyrme model describes nucleons with about $30\%$ error, %
the possibility that it may provide qualitatively correct description of the 
interaction of baryons with gravity can not be excluded. %
It is expected that in the early universe or equivalent high energy experiments, 
the gravitational interaction with baryons is not negligible. %
We hope that our work could provide insight into the observations in such situations. %

\section{Classical gravitating $B=2$ skyrmions}
In this section, we discuss $B=2$ classical regular solutions in the Einstein-Skyrme 
system (ES). %
The Skyrme Lagrangian coupled to gravity is defined by the Lagrangian
\begin{eqnarray}%
	&&\mathcal{L}=\mathcal{L}_G+\mathcal{L}_S ,\\
	&&\mathcal{L}_G =\frac{1}{16\pi G}R ,\label{GLe}\\%
	&&\mathcal{L}_S =\frac{f_{\pi}^2}{16}g^{\mu \nu}{\rm 
Tr}(U^{-1}\partial_{\mu}UU^{-1}\partial_{\nu}U)\nonumber \\%
	&&~~~~+\frac{1}{32e^2}g^{\mu \rho}g^{\nu \sigma} {\rm 
Tr}([U^{-1}\partial_{\mu}U,U^{-1}\partial_{\nu}U]\nonumber \\%
	&&~~~~\times [U^{-1}\partial_{\rho}U,U^{-1}\partial_{\sigma}U]) ,\label{skyLe}%
\end{eqnarray}%
where $f_{\pi}$ is pion decay constant and $e$ is a dimensionless free parameter. %
$U$ describes the $SU(2)$ chiral fields. %
We impose the axially symmetric ansatz on the chiral fields as a possible candidate 
for the $B=2$ minimal energy configuration~\cite{bra}
\begin{eqnarray}%
	U=\cos {F(r,\theta)}+i \bm{\tau}\cdot\bm{n}_R \sin {F(r,\theta)} ,\label{fb2e}%
\end{eqnarray}%
with %
\begin{eqnarray}%
\bm{n}_R=(\sin{\Theta}(r,\theta)\cos{n\varphi},\sin{\Theta}(r,\theta)\sin{n\varphi},\cos{\Theta}(r,\theta)),
\end{eqnarray}%
where $n$ denotes the winding number of solitons, which is equivalent to the baryon number
 $B$. %
Since we are interested in $B=2$ skyrmions, we consider only $n=2$.  %
Correspondingly, the following axially symmetric ansatz is imposed on the 
metric~\cite{kun}%
\begin{eqnarray}%
	ds^2=-fdt^2+\frac{m}{f}(dr^2+r^2d \theta^2)+\frac{l}{f}r^2 \sin^2 \theta d \varphi^2 %
\end{eqnarray}%
where the metric functions $f$ , $m$ and $l$ are the function of coordinates $r$ and $\theta$. %
This metric is symmetric with respect to the $z$-axis ($\theta=0$). %
Substituting these ansatz to the Lagrangian (\ref{skyLe}), %
one obtains the following static energy density for the chiral fields%
\begin{eqnarray}
&&\varepsilon(x,\theta)=  \frac{\sqrt{l}\sin{\theta}}{8}%
	\Bigl \{ x^2 \left( (\partial_{x}F)^2+(\partial_{x}\Theta)^2\sin^2{F} \right) \nonumber 
\\
&&~~~~+(\partial_{\theta}F)^2+(\partial_{\theta}\Theta)^2\sin^2{F}%
	+\frac{n^2m}{l \sin{\theta}}\sin^2{F}\sin^2{\Theta} \Bigr\} \nonumber \\%
&&~~~~+\frac{\sqrt{l}\sin{\theta}}{2} \left[ \frac{f}{m}%
	(\partial_{x}F\partial_{\theta}\Theta-\partial_{\theta}F\partial_{x}\Theta)^2\sin^2{F}\right.
 \nonumber \\%
&&~~~~+\frac{n^2f}{l\sin^2{\theta}}\sin^2{F}\sin^2{\Theta} \Bigl\{ 
\bigl((\partial_{x}F)^2+\frac{1}{x^2}(\partial_{\theta}F)^2\bigr)\nonumber \\%
&&~~~~\left.+\bigl((\partial_{x}\Theta)^2+\frac{1}{x^2}(\partial_{\theta}\Theta)^2\bigr)\sin^2{F}
 \Bigr\} \right]\,, \label{edene}%
\end{eqnarray}
where dimensionless variable $x=e f_{\pi}r$ is introduced. 
The static (classical) energy is thus given by
\begin{eqnarray}
&&M=2\pi \frac{f_{\pi}}{e} \int dxd\theta \varepsilon (x,\theta)\,.\label{cmasse} 
\end{eqnarray}
The covariant topological current is defined by
\begin{eqnarray}%
	B^{\mu}= \frac{\epsilon^{\mu \nu \rho \sigma}}{24 \pi^2} \frac{1}{\sqrt{-g}} %
		 {\rm tr}(U^{-1}\nabla_{\nu}UU^{-1}\nabla_{\rho}UUU^{-1}\nabla_{\sigma}U), \label{bcde} 
\end{eqnarray}%
whose zeroth component corresponds to the baryon number density %
\begin{eqnarray}%
B^0 = -\frac{1}{\pi ^2\sqrt{-g}} \sin^2{F} \sin{\Theta} %
 (\partial_{x} F \partial_{\theta}\Theta-\partial_{\theta} F \partial_{x}\Theta  ). %
\end{eqnarray}%

For the solutions to be regular at the origin $x=0$ and to be asymptotically flat at infinity, %
the following boundary conditions must be imposed %
\begin{eqnarray}%
&&\partial_{x}f(0,\theta)=\partial_{x}m(0,\theta)=\partial_{x}l(0,\theta)=0 , \label{mbr0e} \\%
&&f(\infty ,\theta)=m(\infty ,\theta)=l(\infty ,\theta)=1 . \label{mbr1e}%
\end{eqnarray}%
For the configuration to be axially symmetric, the following boundary conditions must be imposed at $\theta = 0$ and $\pi/2$ %
\begin{eqnarray}%
&&\partial_{\theta}f(x,0)=\partial_{\theta}m(x,0)=\partial_{\theta}l(x,0)=0 , \label{mba0e} \\%
&&\partial_{\theta}f(x,\frac{\pi}{2})=\partial_{\theta}m(x,\frac{\pi}{2})=\partial_{\theta}l(x,\frac{\pi}{2})=0 . \label{mba1e} %
\end{eqnarray}%
For the profile functions, the boundary conditions at the $x=0 , \infty$ are given by%
\begin{eqnarray}%
&&F(0,\theta )= \pi , \ F(\infty,\theta )=0 , \\%
&&\partial_{x}\Theta (0,\theta)= \partial_{x}\Theta (\infty,\theta)=0 . %
\end{eqnarray}%
At $\theta = 0$ and $\pi/2$, %
\begin{eqnarray}%
&&\partial_{\theta}F(x,0)=\partial_{\theta}F(x,\frac{\pi}{2})=0 , \\%
&&\Theta (x,0)=0  \ , \  \Theta (x,\frac{\pi}{2})=\frac{\pi}{2} .%
\end{eqnarray}%
Since the baryon number is defined by the spatial integral of the zeroth component %
of the baryon current, we have %
\begin{eqnarray}%
	B\Eqa{=}\int d^3r \sqrt{-g} B^{0} \nonumber \\%
	\Eqa{=} \frac{1}{2\pi}(2F-\sin{2F})\cos{\Theta} \biggr|_{F_0,\Theta_0}^{F_1,\Theta_1}. %
\end{eqnarray}%
The inner and outer boundary conditions $(F_0,\Theta_0)=(\pi,0)$ and $(F_1,\Theta_1)=(0,\pi)$ yield $B=2$. %

By taking a variation of the static energy (\ref{edene}) with respect to $F$ and $\Theta$, %
one obtains the equations of motion for the profile functions. %
The field equations for the metric functions $f$ , $m$ and $l$ are derived 
from the Einstein equations. We shall show their explicit form in Appendix \ref{feqsa}. 

The effective coupling constant of the Einstein-Skyrme system is given by %
\begin{eqnarray}%
	\alpha = 4 \pi G f_{\pi}^2 
\end{eqnarray}%
which is the only free parameter. %

\begin{center}%
\begin{figure*}%
\begin{tabular}{c@{\hspace{5mm}}c}
\begin{minipage}{80mm}
\begin{center}
\includegraphics[width=7.5cm,keepaspectratio,clip]{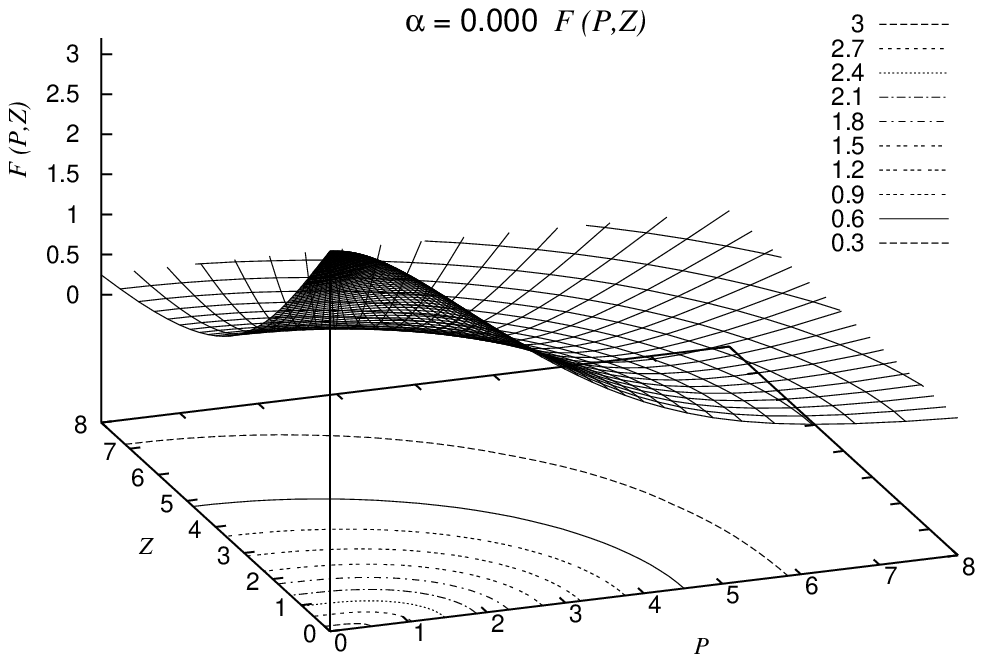}%
\end{center}
\end{minipage}
&
\begin{minipage}{80mm}
\begin{center}
\includegraphics[width=7.5cm,keepaspectratio,clip]{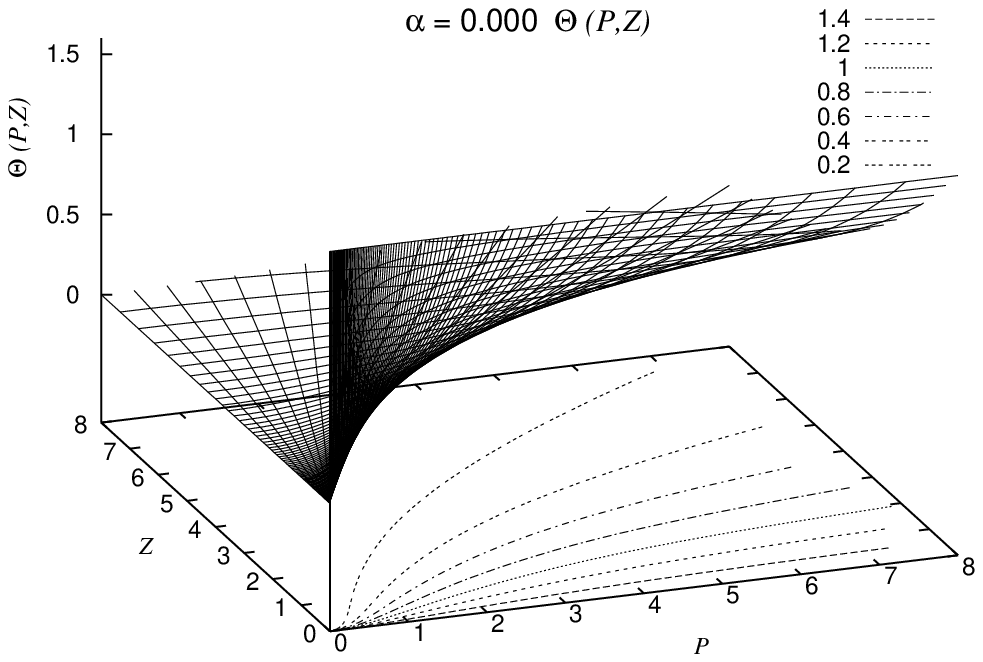}%
\end{center}
\end{minipage}\\

\begin{minipage}{80mm}
\begin{center}
\includegraphics[width=7.5cm,keepaspectratio,clip]{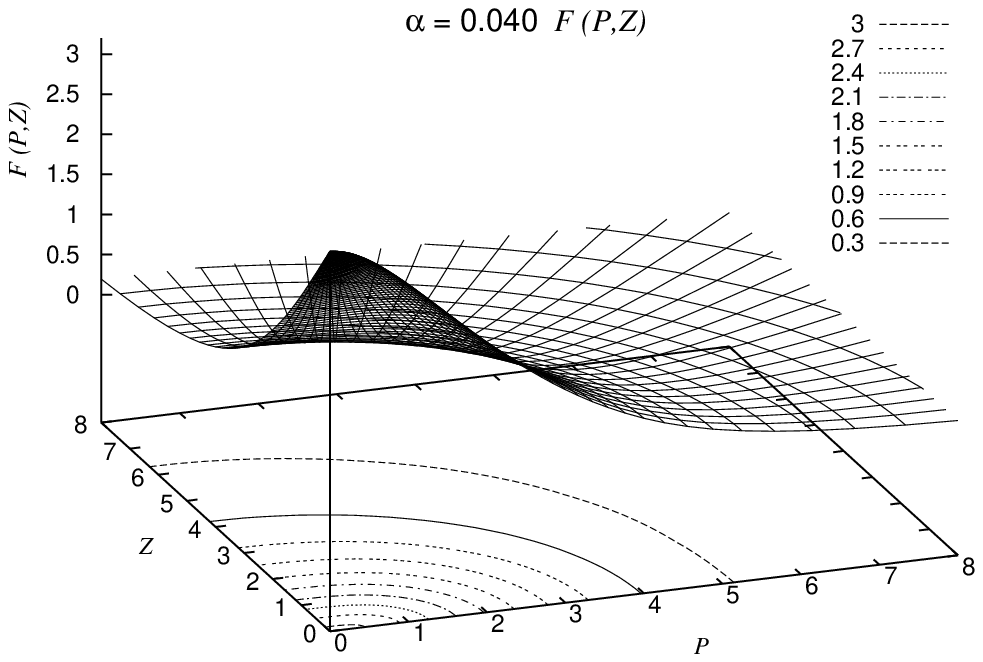}%
\end{center}
\end{minipage}
&
\begin{minipage}{80mm}
\begin{center}
\includegraphics[width=7.5cm,keepaspectratio,clip]{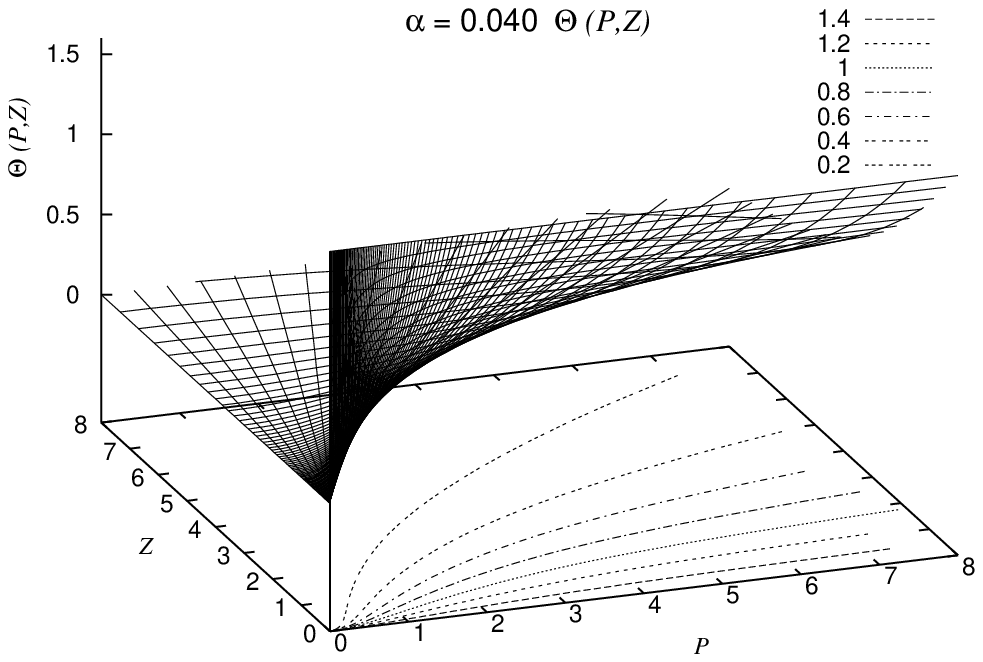}%
\end{center}
\end{minipage}\\

\begin{minipage}{80mm}
\begin{center}
\includegraphics[width=7.5cm,keepaspectratio,clip]{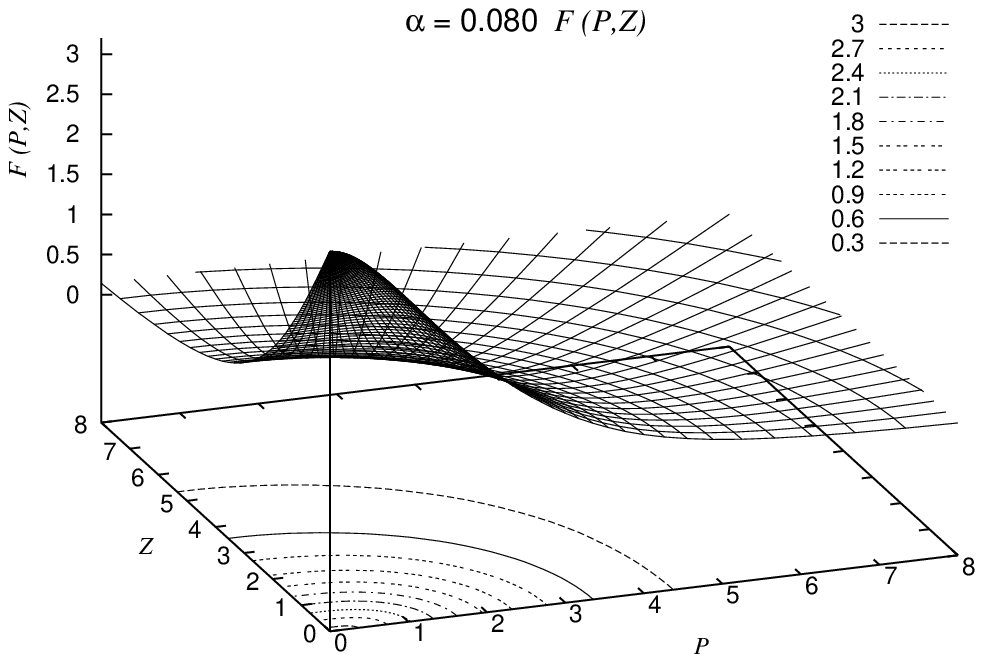}%
\end{center}
\end{minipage}
&
\begin{minipage}{80mm}
\begin{center}
\includegraphics[width=7.5cm,keepaspectratio,clip]{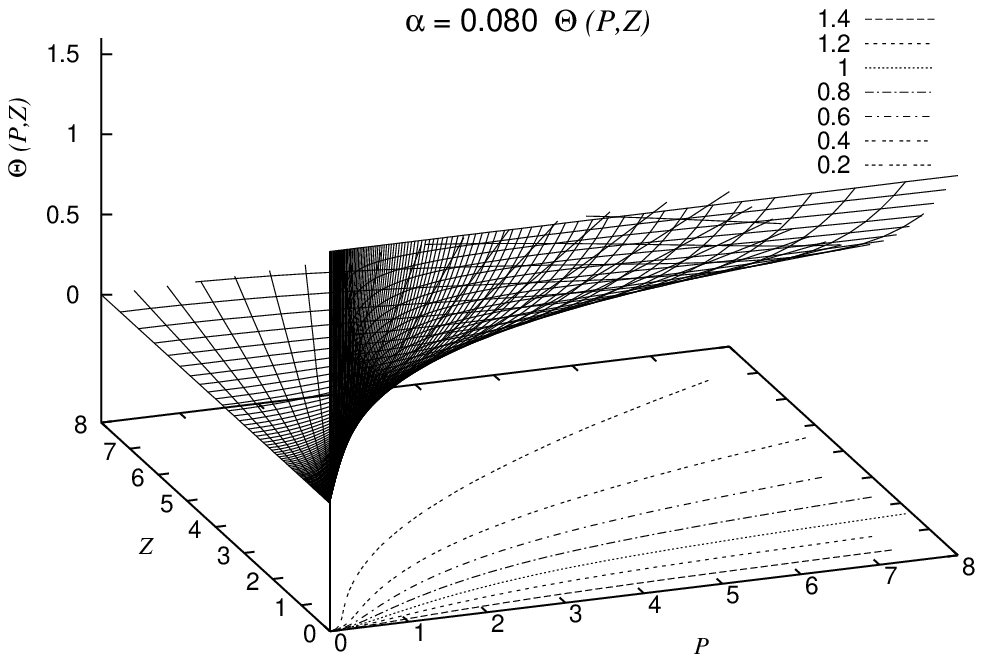}%
\end{center}
\end{minipage}\\

\begin{minipage}{80mm}
\begin{center}
\includegraphics[width=7.5cm,keepaspectratio,clip]{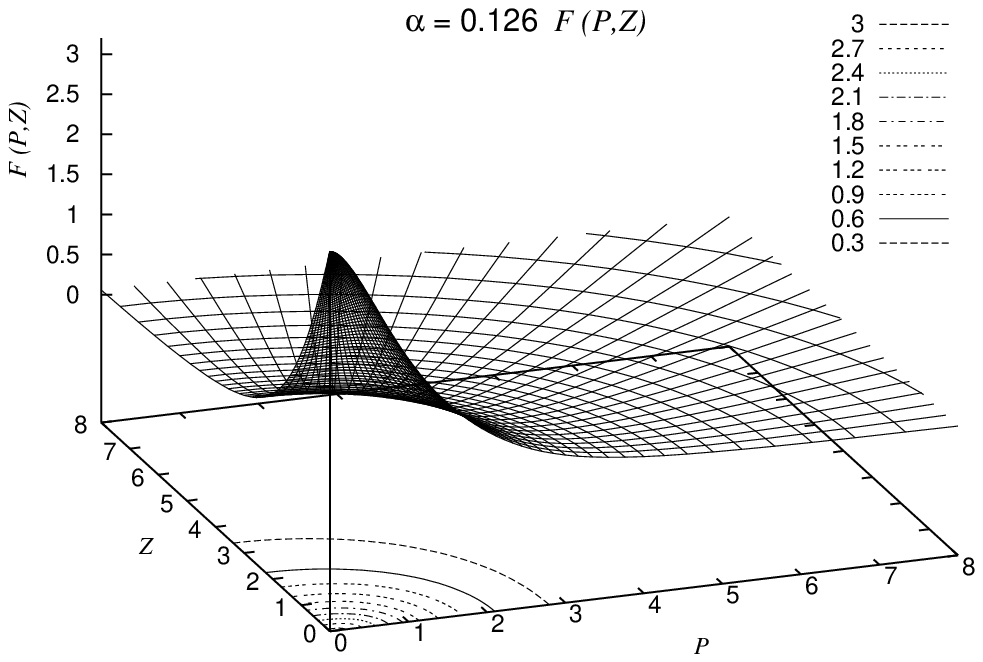}%
\end{center}
\end{minipage}
&
\begin{minipage}{80mm}
\begin{center}
\includegraphics[width=7.5cm,keepaspectratio,clip]{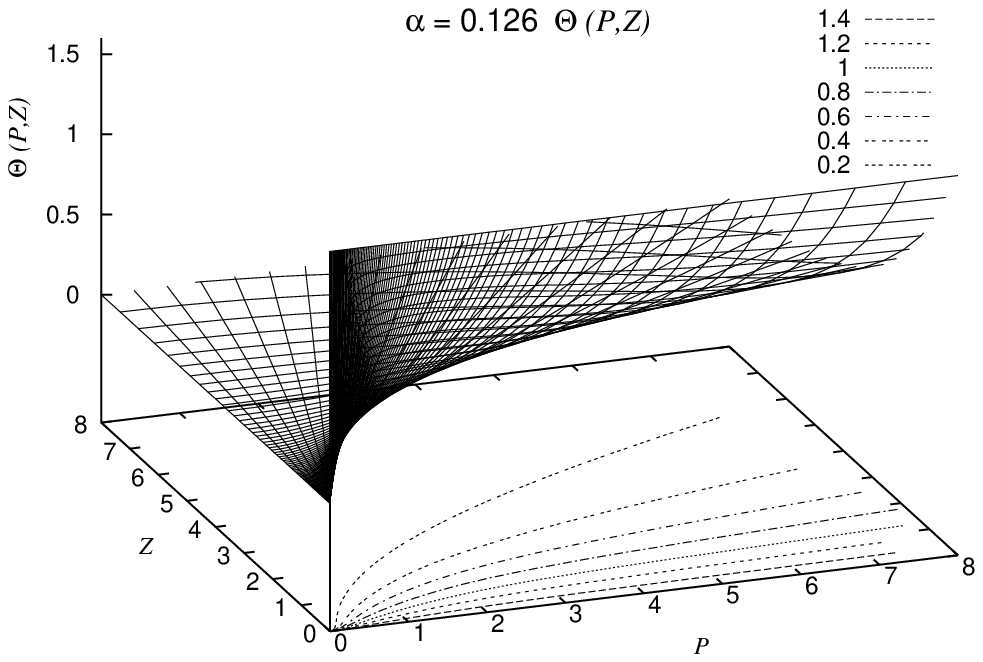}%
\end{center}
\end{minipage}
\end{tabular}
\caption{\label{pfpcf}The profile function $F$ and $\Theta$ 
in the cylindrical coordinate with %
 $\alpha = 0.000 , 0.040 , 0.080 , 0.126$ are shown.
We use dimensionless variables $P=ef_{\pi}\rho$ and $Z=ef_{\pi}z$. %
There exists no solution for $\alpha \gtrsim 0.127$. }%
\end{figure*}%
\end{center}

\begin{figure}%
\begin{minipage}{80mm}%
\begin{center}
\includegraphics[width=6.5cm,keepaspectratio,clip]{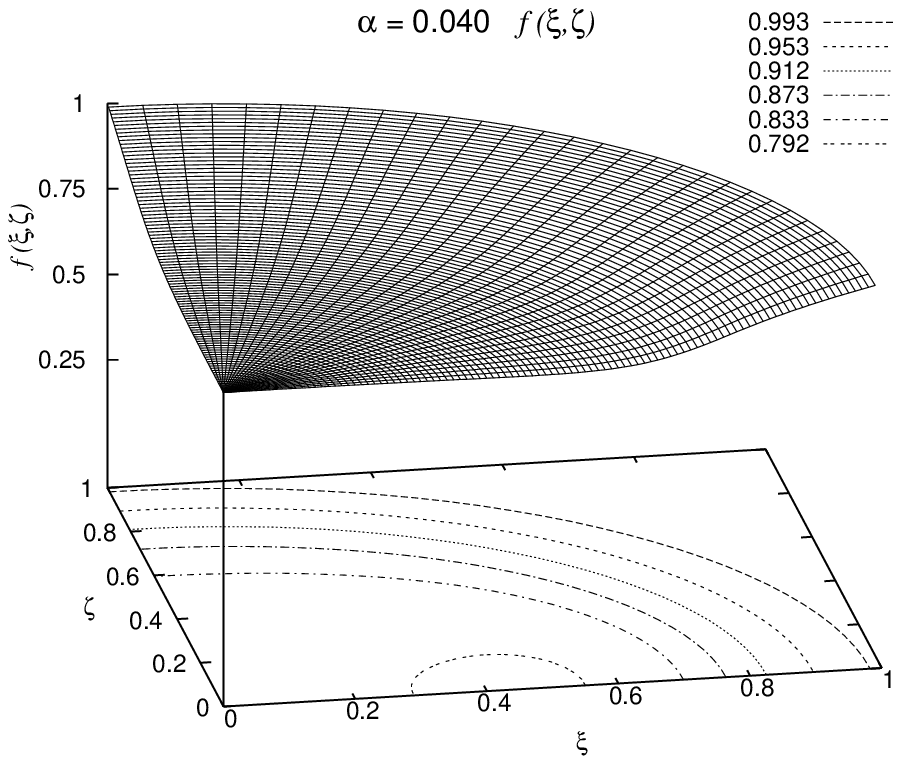}%
\vspace{3mm}
\includegraphics[width=6.5cm,keepaspectratio,clip]{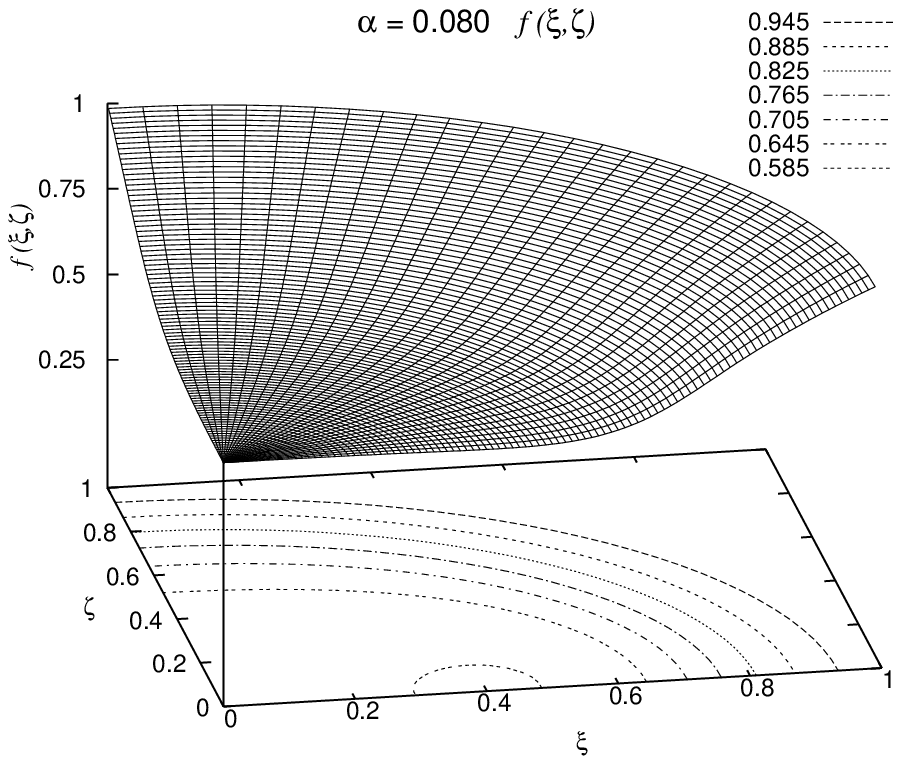}%
\vspace{3mm}
\includegraphics[width=6.5cm,keepaspectratio,clip]{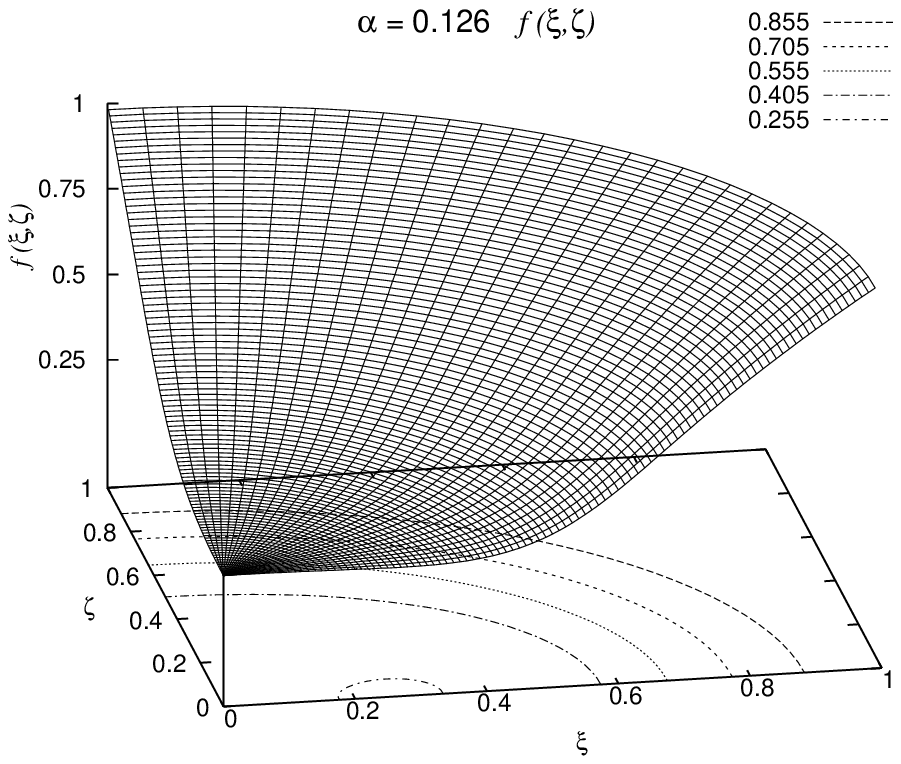}%
\end{center}
\end{minipage}
\caption{\label{mff}The metric function $f$ in the cylindrical coordinate 
with $\alpha = 0.040 , 0.080 , 0.126 $ is shown. %
We use $\xi=P/(1+P)$ and $\zeta = Z/(1+Z)$, %
instead of dimensionless variables $P=ef_{\pi}\rho$ and $Z=ef_{\pi}z$. %
}%
\end{figure}

\begin{figure}%
\begin{minipage}{80mm}
\begin{center}
\includegraphics[width=6.5cm,keepaspectratio,clip]{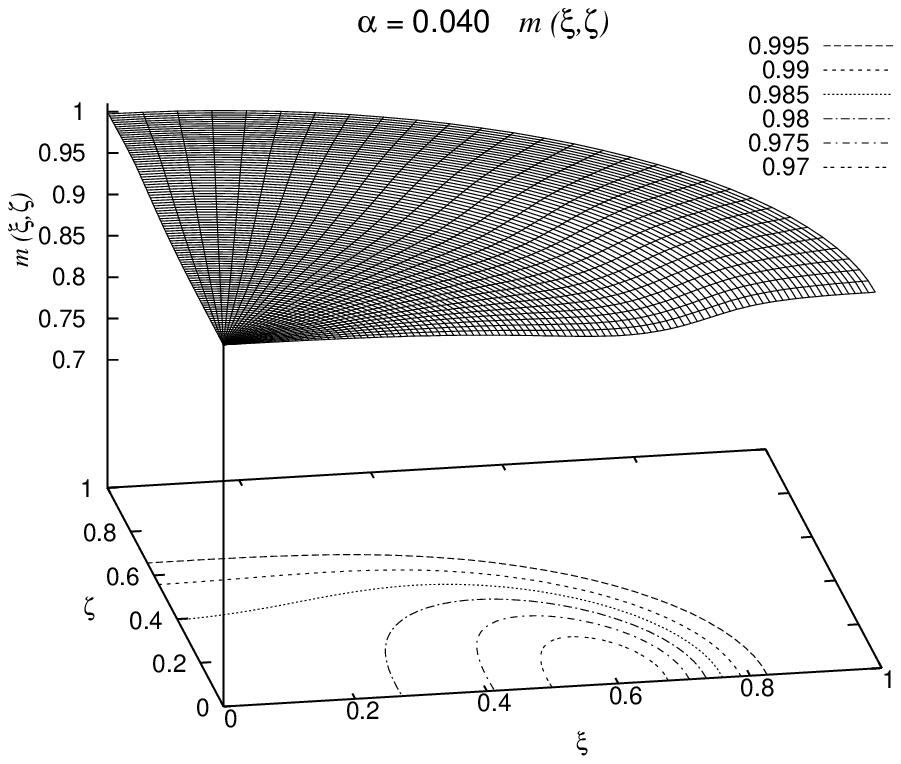}%
\vspace{3mm}
\includegraphics[width=6.5cm,keepaspectratio,clip]{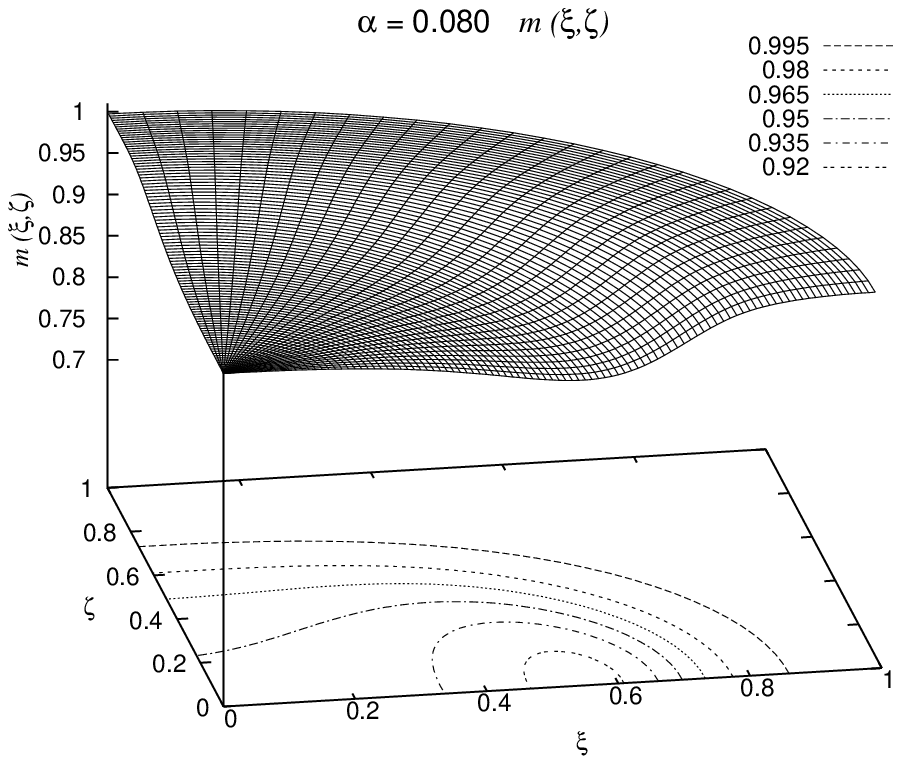}%
\vspace{3mm}
\includegraphics[width=6.5cm,keepaspectratio,clip]{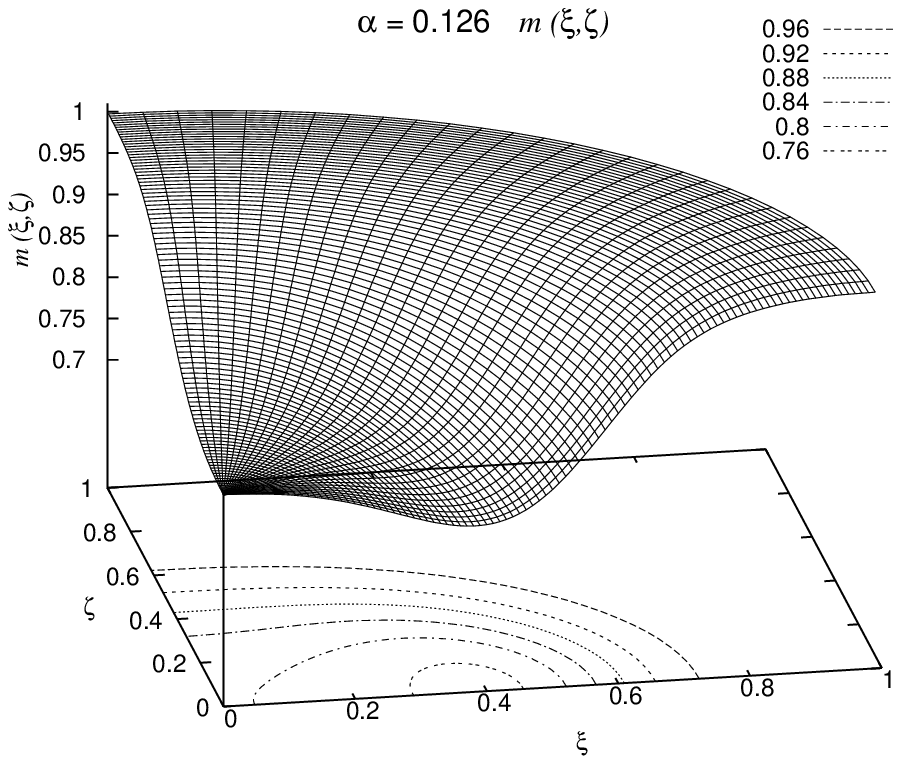}%
\end{center}
\end{minipage}
\caption{\label{mmf}Same as Fig.\ref{mff} for the metric function $m$.}%
\end{figure}

\begin{figure}%
\begin{tabular}{c}%
\begin{minipage}{80mm}%
\begin{center}
\includegraphics[width=6.5cm,keepaspectratio,clip]{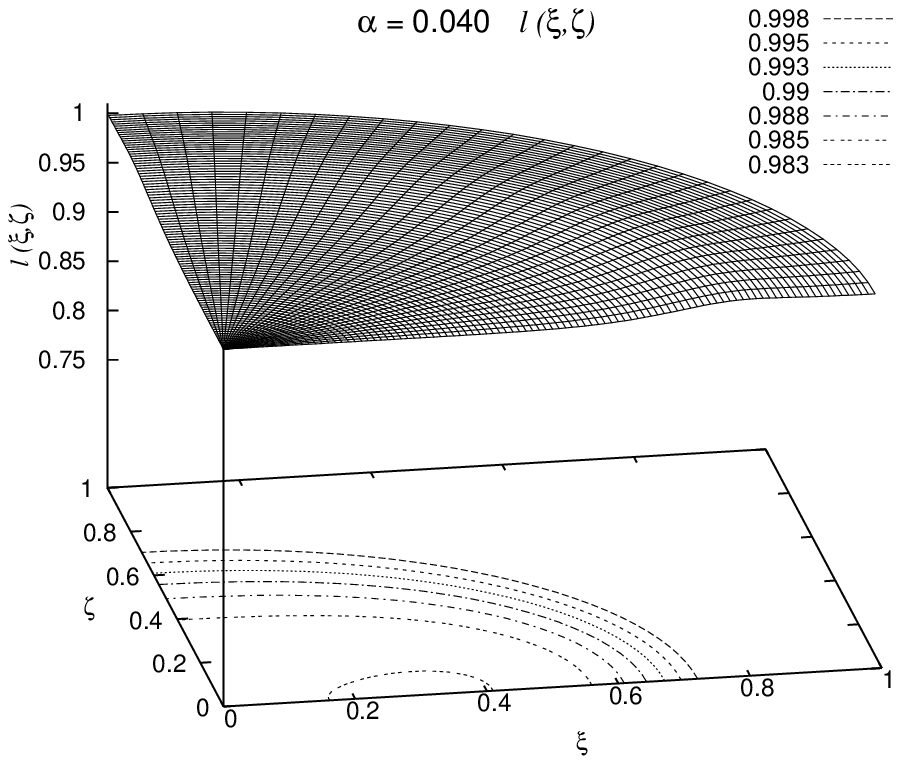}%
\vspace{3mm}
\includegraphics[width=6.5cm,keepaspectratio,clip]{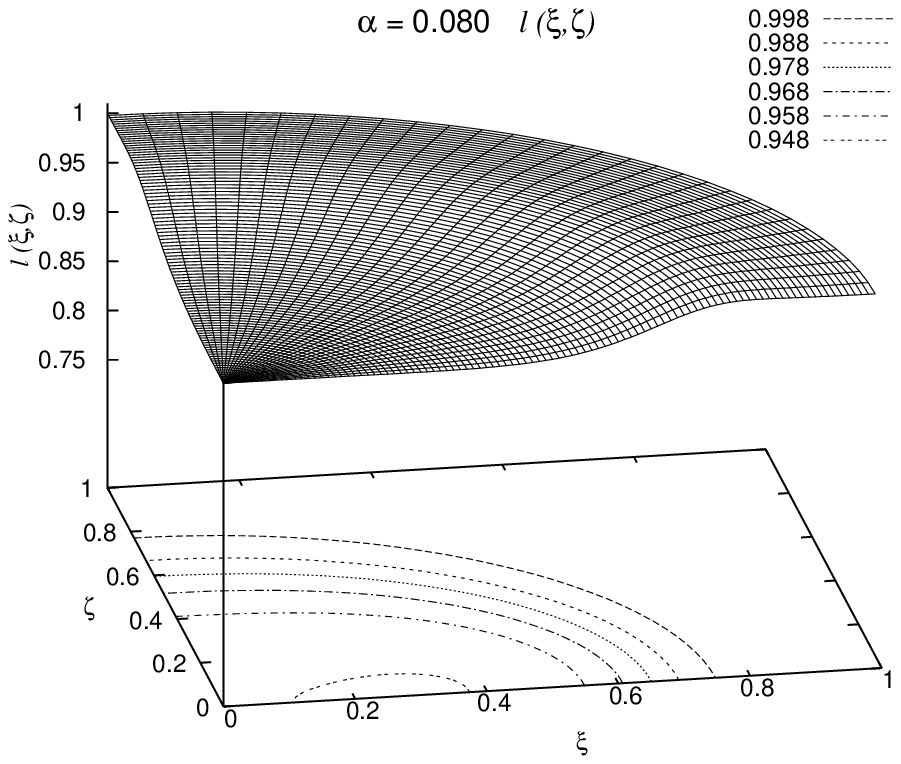}%
\vspace{3mm}
\includegraphics[width=6.5cm,keepaspectratio,clip]{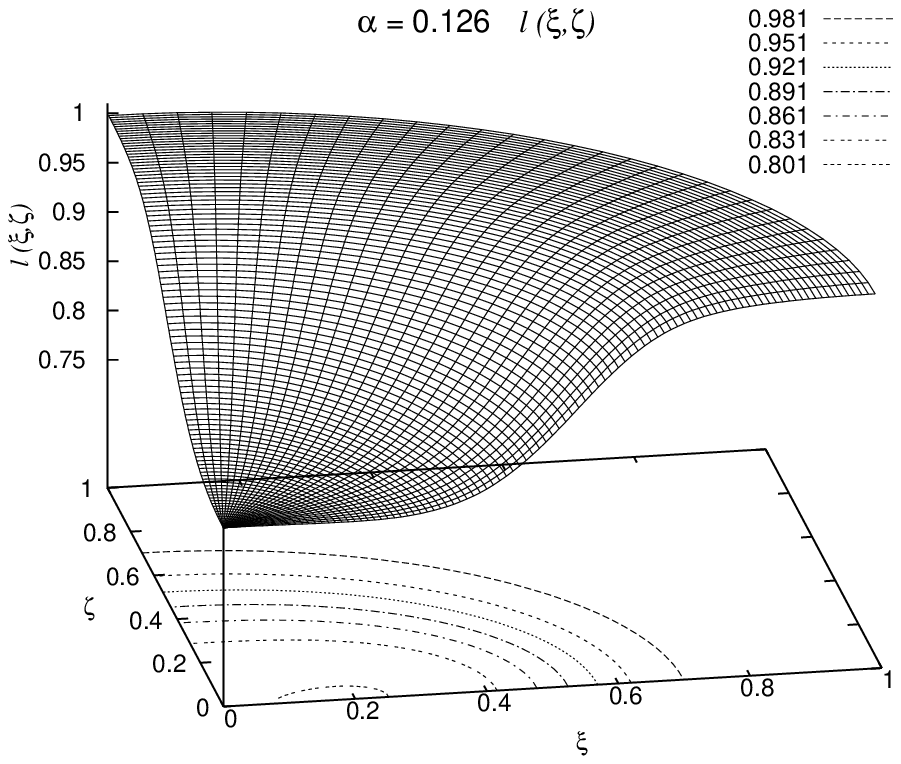}%
\vspace{3mm}
\end{center}
\caption{\label{mlf}Same as Fig.\ref{mff} for  the metric function $l$.}%
\end{minipage}
\end{tabular}%
\end{figure}

\begin{figure}%
\begin{tabular}{c}%
\begin{minipage}{80mm}
\begin{center}
\includegraphics[width=7.5cm,keepaspectratio,clip]{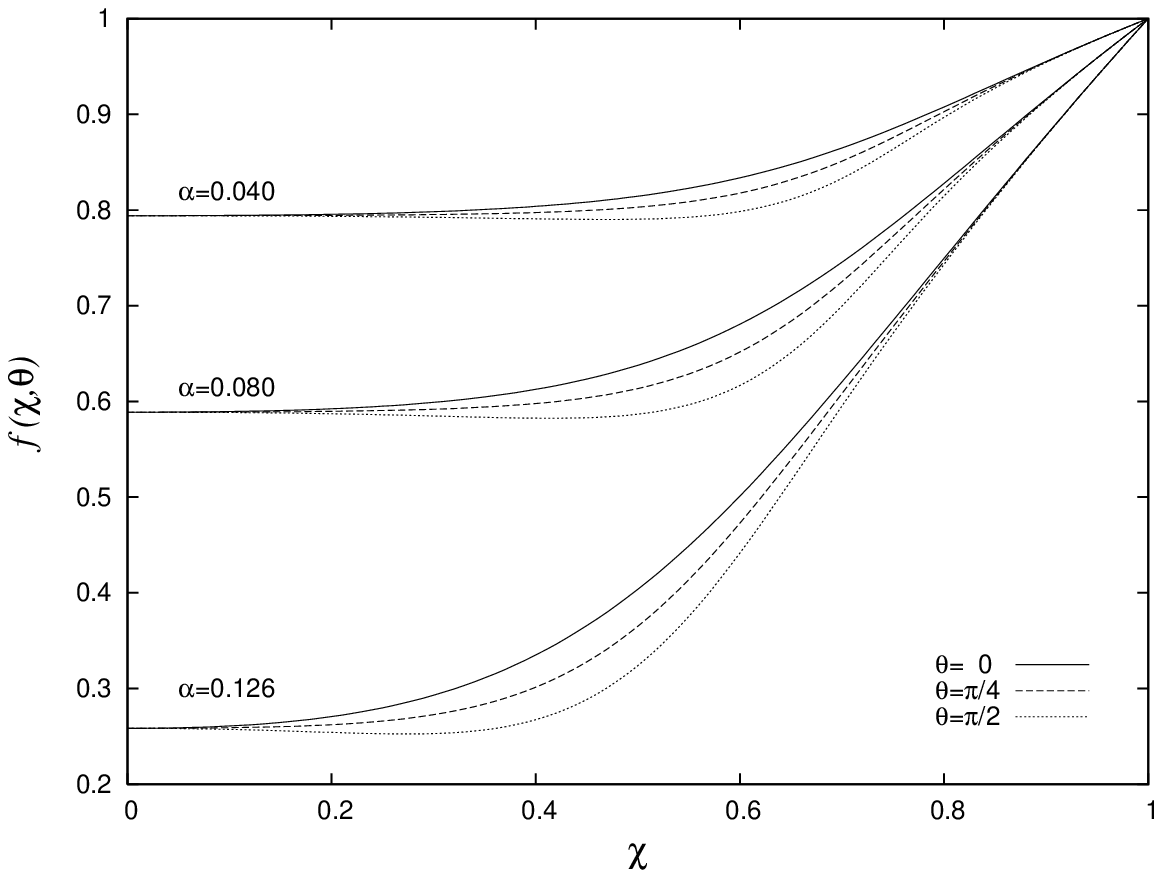}%
\vspace{3mm}
\includegraphics[width=7.5cm,keepaspectratio,clip]{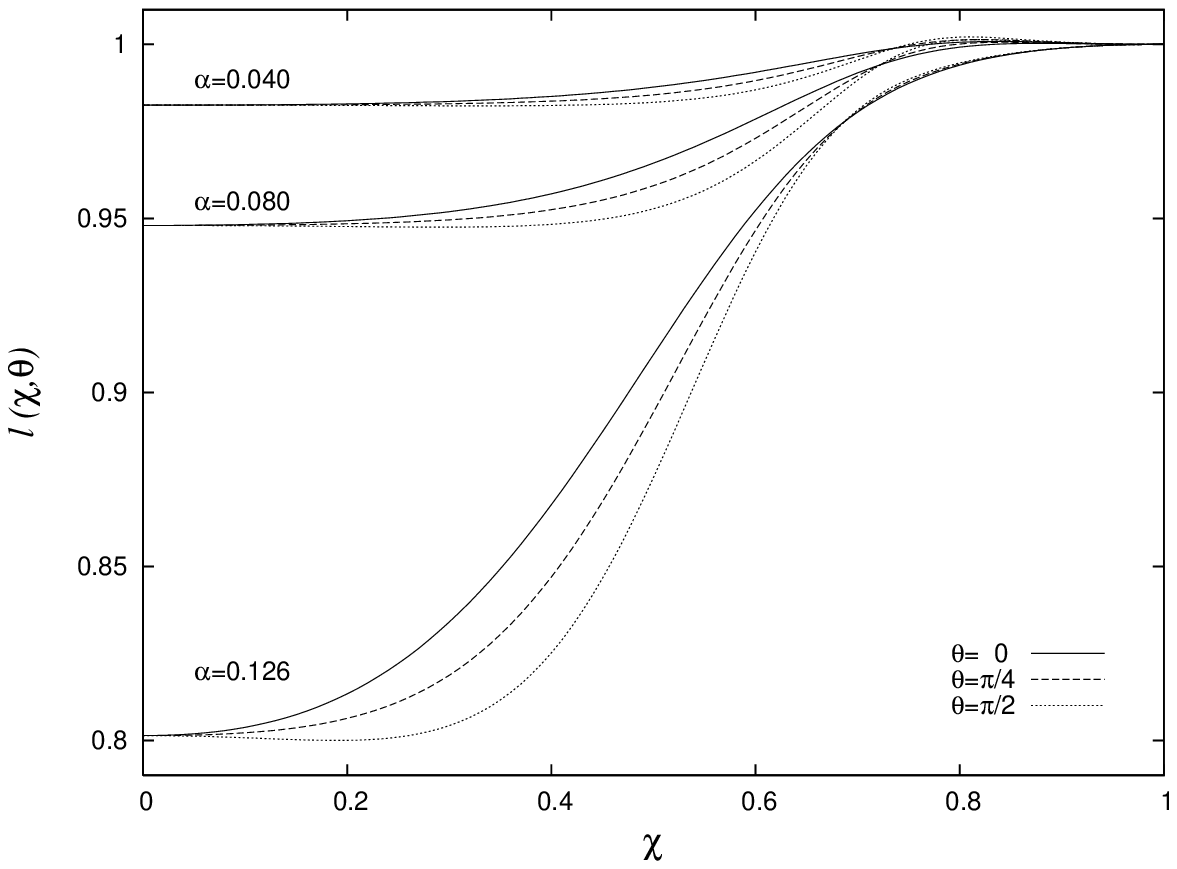}%
\vspace{3mm}
\includegraphics[width=7.5cm,keepaspectratio,clip]{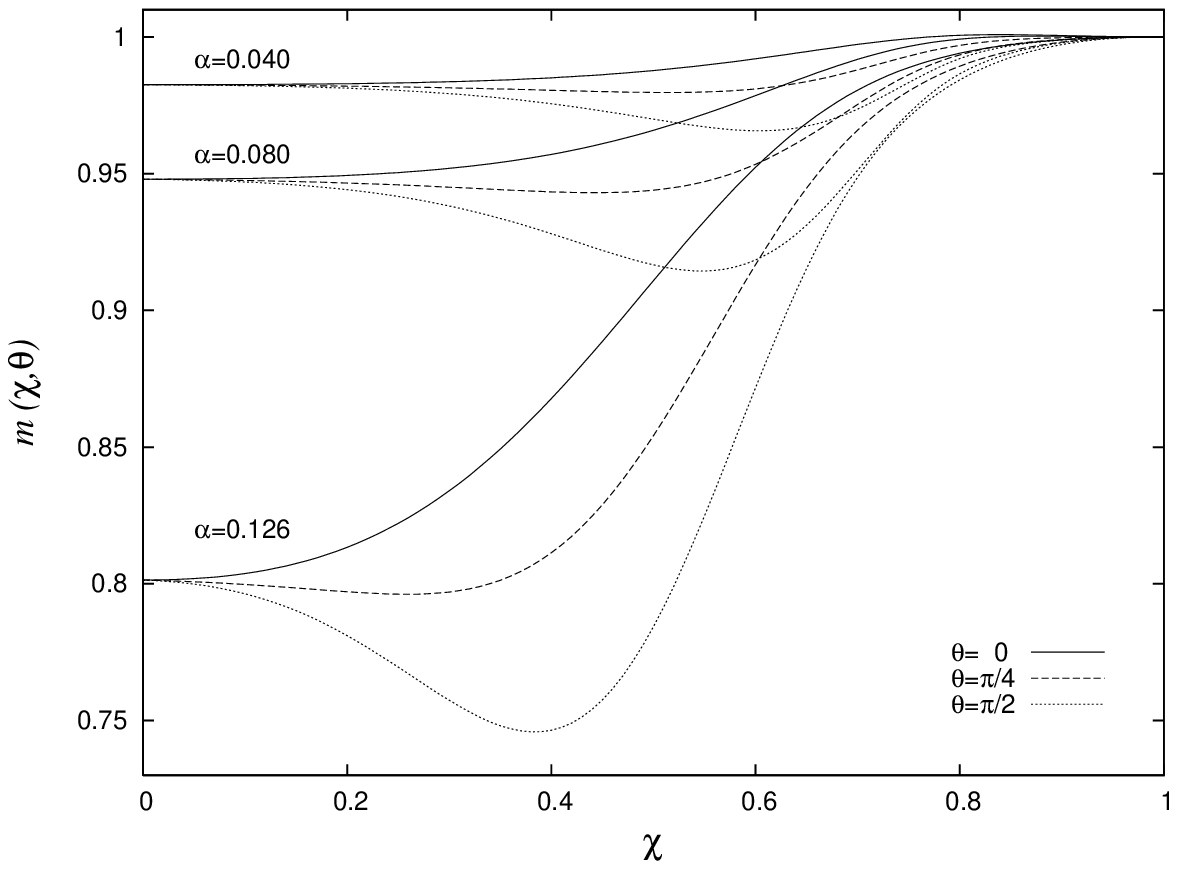}%
\end{center}
\caption{\label{secf} The metric function $f$, $m$ and $l$ 
for $\theta = 0 , \pi/4 , \pi/2$ as a function of radial 
coordinate with  $\alpha = 0.040 , 0.080 , 0.126$ are shown. %
We use $\chi=x/(1+x)$, instead of dimensionless 
variable $x=e f_{\pi}r$.}%
\end{minipage}
\end{tabular}%
\end{figure}

\begin{figure}%
\includegraphics[width=7.5cm,keepaspectratio,clip]{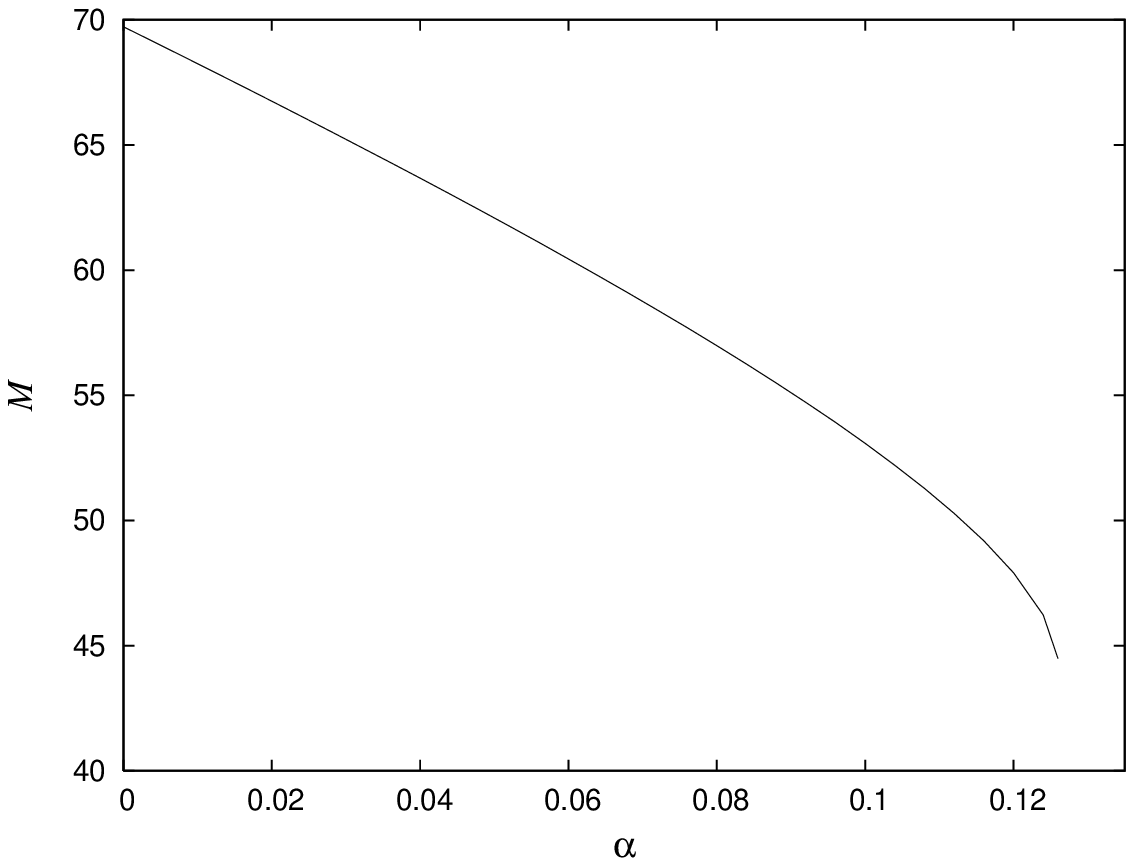}%
\caption{\label{cmassf} The coupling constant dependence of the \
classical $B=2$ skyrmion dimensionless mass derived from 
Eq.(\ref{cmasse}). %
}%
\end{figure}

We use the relaxation method to solve these nonlinear equations with the typical grid size $100 \times 30$. %
In Fig.\ref{pfpcf}, the profile functions for $\alpha=0,0.04,0.08,0.126$ is presented. %
No solution exist for $\alpha \gtrsim 0.127$. %
Also, the metric functions are shown in Figs.\ref{mff}$-$\ref{mlf}. %
In Fig.\ref{secf}, we display the metric functions at $\theta=0,\pi /4,\pi /2$ as %
a function of radial coordinate. %
Fig.\ref{cmassf} shows $\alpha$ dependence of the static energy $M$ in unit of $f_{\pi}/e$. %
In Fig. \ref{edenf} the energy densities defined in Eq.(\ref{edene}) are plotted. %
The baryon densities $b=\sqrt{-g}B^0$ are shown in Fig. \ref{bdenf}. 



\section{Collective quantization scheme for the $B=2$ soliton solution}

\begin{center}%
\begin{figure*}%
\begin{tabular}{c@{\hspace{5mm}}c}

\begin{minipage}{80mm}
\begin{center}%
\includegraphics[width=7.5cm,keepaspectratio,clip]{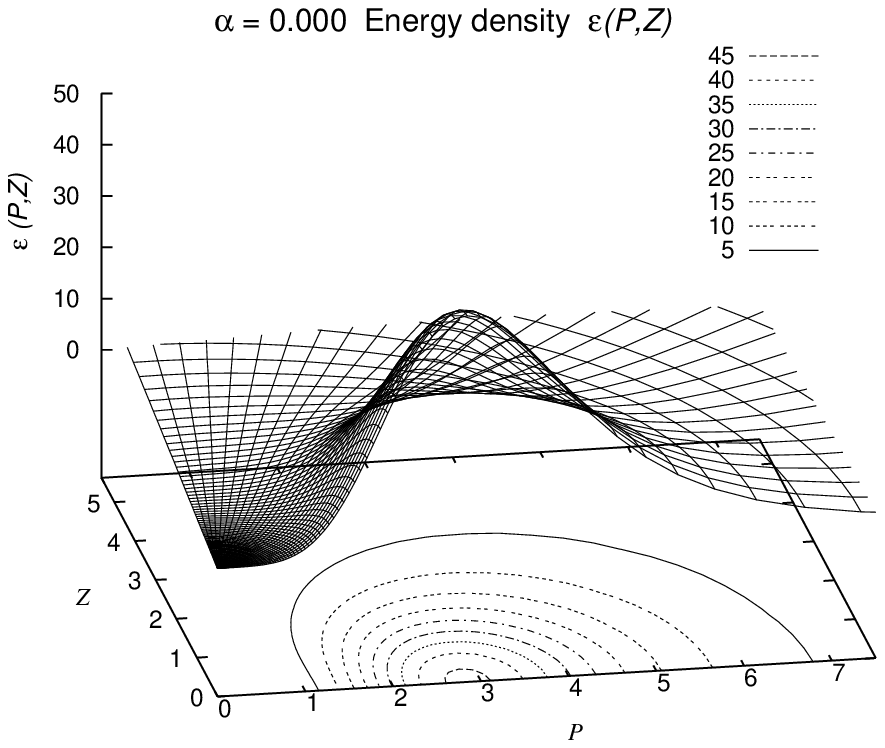}%
\end{center}%
\end{minipage}%
&
\begin{minipage}{80mm}
\begin{center}
\includegraphics[width=7.5cm,keepaspectratio,clip]{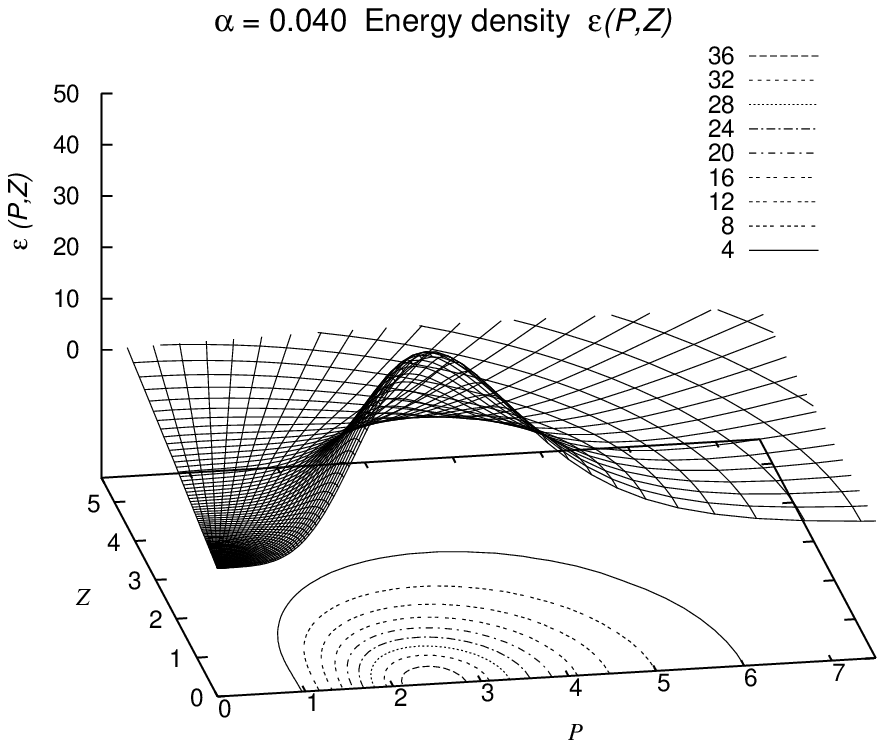}%
\end{center}
\end{minipage}\\

\begin{minipage}{80mm}
\begin{center}
\includegraphics[width=7.5cm,keepaspectratio,clip]{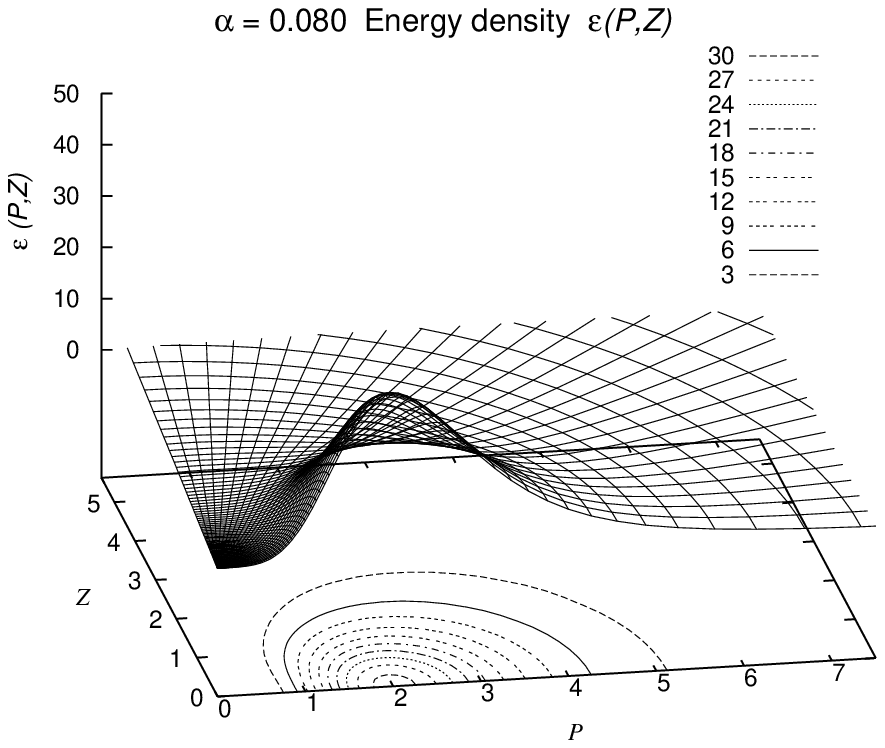}%
\end{center}
\end{minipage}
&
\begin{minipage}{80mm}
\begin{center}
\includegraphics[width=7.5cm,keepaspectratio,clip]{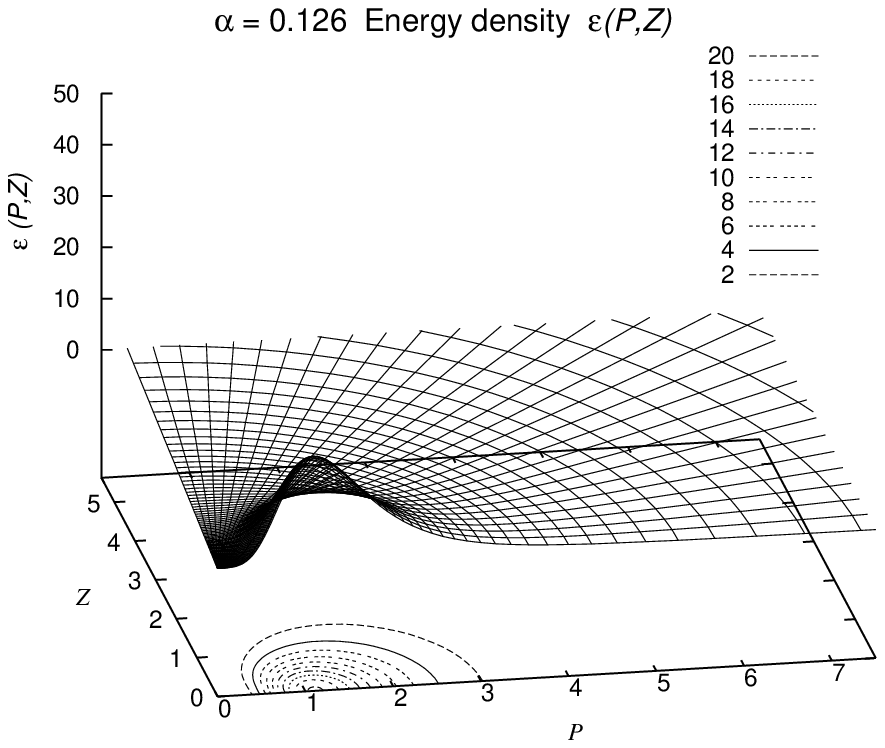}%
\end{center}
\end{minipage}
\end{tabular}
\caption{\label{edenf}
The energy density $\varepsilon$  in the cylindrical coordinate
with $\alpha = 0.000 , 0.040 , 0.080 , 0.126$ are shown, 
in terms of the dimensionless variables $P=ef_{\pi}\rho$ and $Z=ef_{\pi}z$. %
}%
\end{figure*}%
\end{center}%
\begin{center}%
\begin{figure*}%
\begin{tabular}{c@{\hspace{5mm}}c}
\begin{minipage}{80mm}
\begin{center}%
\includegraphics[width=7.5cm,keepaspectratio,clip]{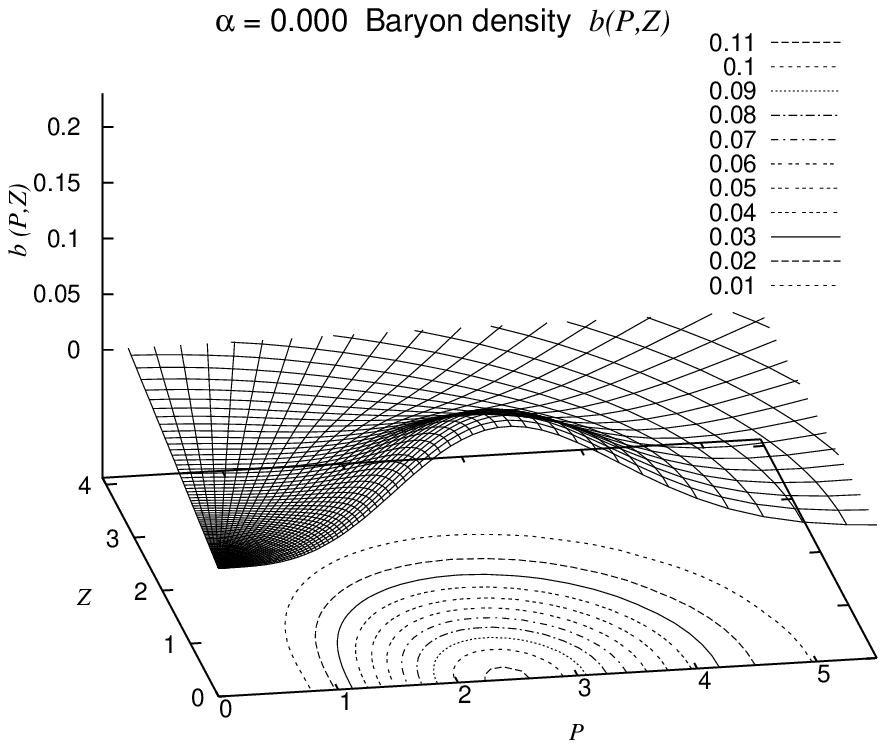}%
\end{center}%
\end{minipage}%
&
\begin{minipage}{80mm}
\begin{center}
\includegraphics[width=7.5cm,keepaspectratio,clip]{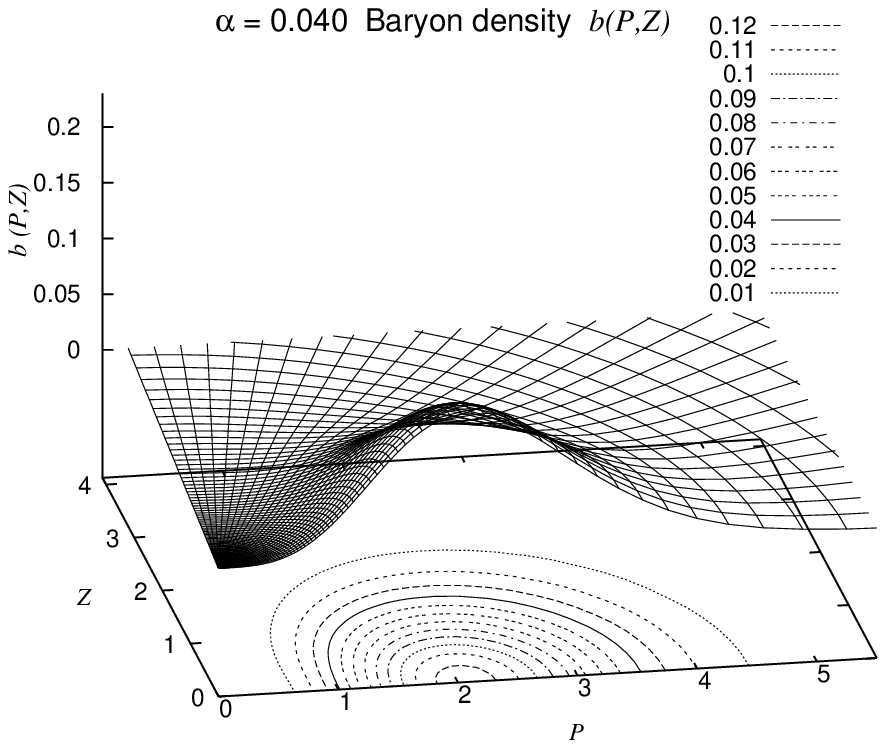}%
\end{center}
\end{minipage}\\

\begin{minipage}{80mm}
\begin{center}
\includegraphics[width=7.5cm,keepaspectratio,clip]{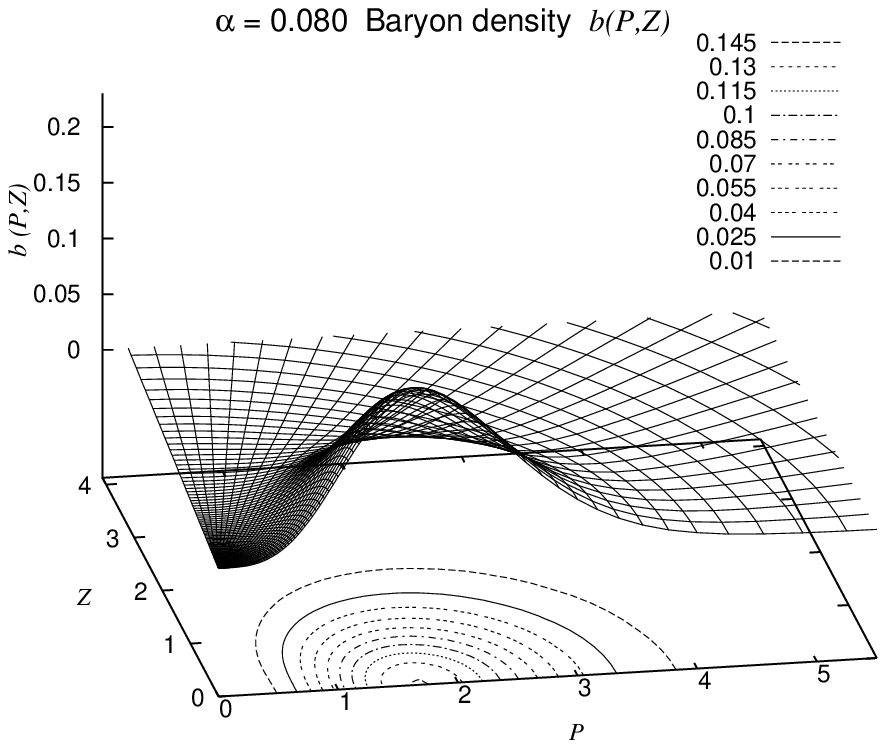}%
\end{center}
\end{minipage}
&
\begin{minipage}{80mm}
\begin{center}
\includegraphics[width=7.5cm,keepaspectratio,clip]{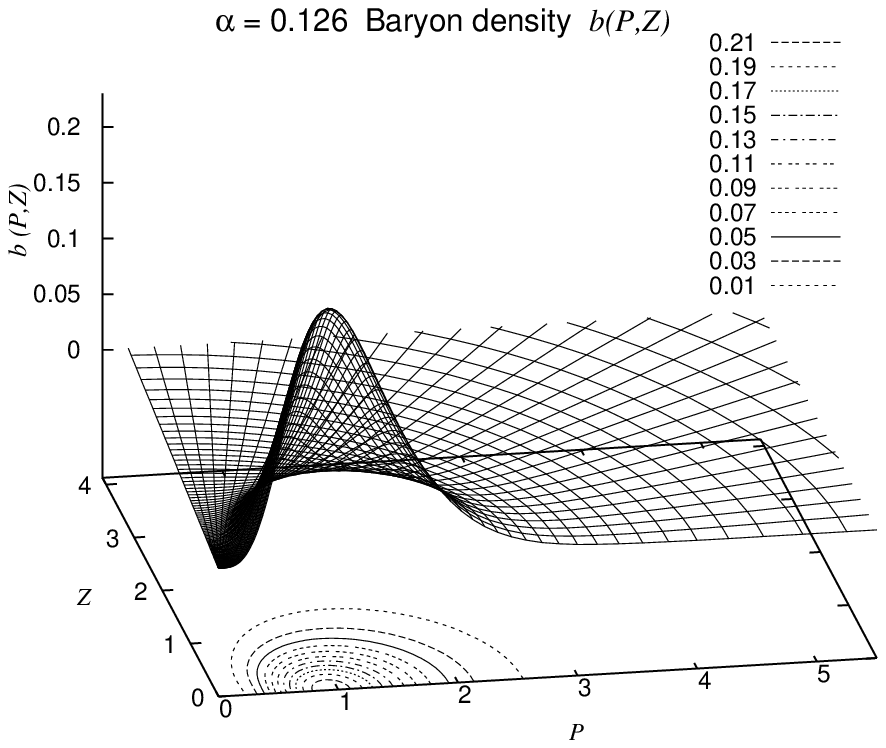}%
\end{center}
\end{minipage}

\end{tabular}
\caption{\label{bdenf} \\
The baryon density $b=\sqrt{-g}B^0$ in the cylindrical coordinate 
with $\alpha = 0.000 , 0.040 , 0.080 , 0.126$ are shown, 
in terms of the dimensionless variables $P=ef_{\pi}\rho$ and $Z=ef_{\pi}z$. %
}%
\end{figure*}%
\end{center}%

In this section, let us try to give an interpretation to the $B=2$ skyrmion solution 
obtained in the last section as a baryonic object by assigning quantum number of 
spin and isospin to it. 
We employ the semi-classical zero-mode quantization following Ref.~\cite{bra}.
Let us introduce dynamically rotated chiral fields
\begin{eqnarray}%
\hat{U}(\bm{r},t)=A(t)U(\bm{r}')A^{\dagger}(t),~~
\bm{r}'=R(B(t))\bm{r}, \label{rote}%
\end{eqnarray}%
and 
\begin{eqnarray}%
R_{ij}(B)=\frac{1}{2}{\rm Tr}(\bm{\tau}_i B \bm{\tau}_j B^{\dagger}) ,\label{rotme}%
\end{eqnarray}%
where $A(t)$ and $B(t)$ are the time dependent $SU(2)$ matrices generating the isospin and the spatial rotations. %
Substituting (\ref{rote}) into the Lagrangian (\ref{skyLe}), one finds
\begin{eqnarray}%
L=-M+\frac{1}{2}a_{i}U_{ij}a_{j}-a_{i}W_{ij}b_{j}+\frac{1}{2}b_{i}V_{ij}b_{j}, 
\label{qlage}%
\end{eqnarray}
where $M$ is the static energy of the $B=2$ skyrmion given in Eq.(\ref{cmasse}).
The Lagrangian is quadratic in time derivatives %
\begin{eqnarray}%
a_i=-i{\rm Tr}\tau_{i}A^{\dagger}\dot{A} \ , \ b_{i}=i{\rm Tr}\tau_{i}\dot{B}B^{\dagger}.\label{tope}%
\end{eqnarray}%
The moments of inertia tensor $U_{ij}$ , $V_{ij}$ and $W_{ij}$ are expressed in terms of the chiral field $U$ as 
\begin{eqnarray}%
U_{ij}\Eqa{=} {\textstyle \frac{1}{8f_{\pi}e^3}}\int d^3x \Bigl[ {\textstyle\frac{m\sqrt{l}}{f^2}}x^2 \sin{\theta} %
			\bigl\{ {\rm Tr}\bigl(U^{\dagger}[{\textstyle \frac{\bm{\tau}_i}{2}},U]%
						     U^{\dagger}\bigl[{\textstyle \frac{\bm{\tau}_j}{2}},U\bigr] \bigr) \nonumber \\%
\Eqa{+}g^{kl}{\rm Tr} \bigl[U^{\dagger}\partial_k U,U^{\dagger} [{\textstyle \frac{\bm{\tau}_i}{2}},U] \bigr] %
	\bigl[U^{\dagger}\partial_l U,U^{\dagger}[{\textstyle \frac{\bm{\tau}_j}{2}},U] \bigr] \bigr\} \Bigr], \label{moiue}\\%
W_{ij}\Eqa{=}U_{ij} \left\{ [{\textstyle \frac{\bm{\tau}_j}{2}},U] \rightarrow i(\bm{x}' times \nabla)_jU ) \right\}, \label{moiwe} \\ %
V_{ij}\Eqa{=}W_{ij} \left\{ [{\textstyle \frac{\bm{\tau}_i}{2}},U] \rightarrow i(\bm{x}' times \nabla)_iU ) \right\}. \label{moive}%
\end{eqnarray}%
Body-fixed isospin operator $K_i$ and angular momentum operator $L_i$ %
are defined as a canonically conjugate to $a_i$ and $b_i$ %
\begin{eqnarray}%
K_i\Eqa{=}\frac{\partial L}{\partial a_i}=U_{ij}a_j-W_{ij}b_j , \nonumber \\%
L_i\Eqa{=}\frac{\partial L}{\partial b_i}=-W_{ij}^{T}a_j+V_{ij}b_j . \label{bfoe}%
\end{eqnarray}%
These operators are related to the usual coordinate-fixed isospin $I_i$ and spin $J_i$ by the orthogonal transformation, %
\begin{eqnarray}%
I_i\Eqa{=}-R_{ij}(A)K_j \ , \ J_i=-R_{ij}(B)^{T}L_j. \label{cfoe} %
\end{eqnarray}%
The commutation relations for these operators are
\begin{eqnarray}%
&&[ K_i,K_j ] = i \epsilon_{ijk}K_k \ , \ [ L_i,L_j ] = i \epsilon_{ijk}L_k, \nonumber \\%
&&[ I_i,I_j ] = i \epsilon_{ijk}I_k \ , \ [ J_i,J_j ] = i \epsilon_{ijk}J_k.  %
\end{eqnarray}%
These operators satisfy the Casimir invariance by using Eq.(\ref{cfoe}) %
\begin{eqnarray}%
\bm{K}^2\Eqa{=}\bm{I}^2 \ , \ \bm{L}^2=\bm{J}^2 .\label{casie}%
\end{eqnarray}%
The symmetry of the classical soliton induces the following conditions for the inertia tensors
\begin{eqnarray}%
&&U_{11}=U_{22} \ , \ V_{11}=V_{22} \ , \ W_{11}=W_{22} =0 ,\nonumber \\%
&&~~~~~~V_{33}=4U_{33} \ , \ W_{33}=2U_{33}. \label{midie}%
\end{eqnarray}%
Thus it is sufficient to calculate only $U_{11}$ , $U_{33}$ and $V_{11}$. %
Inserting (\ref{fb2e}) into Eqs.(\ref{moiue}) and (\ref{moive}) one obtains %
\begin{eqnarray}%
U_{11}\Eqa{=}{\textstyle \frac{\pi}{f_{\pi}e^3}} \int dx d\theta \Bigl[ {\textstyle \frac{m\sqrt{l}}{4f^2}}x^2 \sin{\theta} %
			\sin^2{F}(1+\cos^2{\Theta}) \nonumber \\ %
&& \hspace{3mm}+{\textstyle \frac{\sqrt{l}}{f}}\sin{\theta}\sin^2{F} \bigl\{(1+\cos^2{\Theta})(x^2F_{,x}^2+F_{,\theta}^2) \nonumber \\ %
&& \hspace{3mm}+\sin^2{F}\cos^2{\Theta}(x^2\Theta_{,x}^2+\Theta_{,\theta}^2) \nonumber \\ %
&& \hspace{3mm}+{\textstyle \frac{n^2m}{l \sin^2{\theta}}}\sin^2{F}\sin^2{\Theta} \bigr \} \Bigr], \label{u11e} \\%
U_{33}\Eqa{=}{\textstyle \frac{\pi}{f_{\pi}e^3}} \int dx d\theta %
			\Bigl[ {\textstyle \frac{m\sqrt{l}}{2f^2}}x^2\sin{\theta} %
			\sin^2{F}\sin^2{\Theta} \nonumber \\ %
&& \hspace{3mm}+{\textstyle \frac{2\sqrt{l}}{f}}\sin{\theta}\sin^2{F}\sin^2{\Theta} %
	 \bigl\{x^2F_{,x}^2+F_{,\theta}^2 \nonumber \\ %
&& \hspace{13mm}+\sin^2{F}(x^2\Theta_{,x}^2+\Theta_{,\theta}^2) %
	\bigr \} \Bigr], \label{u33e} \\ %
V_{11}\Eqa{=}{\textstyle \frac{\pi}{f_{\pi}e^3}} \int dx d\theta %
			\Bigl[ {\textstyle \frac{m\sqrt{l}}{4f^2}}x^2\sin{\theta} %
			(F_{,x}^2+\Theta_{,\theta}^2\sin^2{F} \nonumber \\ %
&& \hspace{33mm}  +n^2\cot^2{\theta}\sin^2{F}\sin^2{\Theta}) \nonumber \\ %
&& \hspace{3mm}+{\textstyle \frac{\sqrt{l}}{f}}x^2\sin{\theta}\sin^2{F} %
	\bigl\{(F_{,x}\Theta_{,\theta}-F_{,\theta}\Theta_{,x})^2 \nonumber \\ %
&& \hspace{18mm}+n^2(F_{,x}^2+\Theta_{,x}^2\sin^2{F})\cot^2{\theta}\sin^2{\Theta} \bigr \} \nonumber \\ %
&& \hspace{3mm}+{\textstyle \frac{n^2 \sqrt{l}}{f \sin{\theta}}}(\cos^2{\theta}+{\textstyle \frac{m}{l}}) \nonumber \\%
&& \hspace{13mm}\times (F_{,\theta}^2+\Theta_{,\theta}^2\sin^2{F}) \sin^2{F} \sin^2{\Theta} \Bigr]. \label{v11e}%
\end{eqnarray}%
From Eqs.(\ref{bfoe}) and (\ref{midie}) we derive the constraint %
\begin{eqnarray}%
(2K_3+L_3)\left| phys \right>=0. \label{conste}%
\end{eqnarray}%
Inserting the body-fixed operator (\ref{bfoe}), Casimir invariant (\ref{casie}) and 
constraint (\ref{conste}) into the Lagrangian (\ref{qlage}) one can get the Hamiltonian operator as 
\begin{eqnarray}%
	H=M\Eqa{+}\frac{\bm{I}^2}{2U_{11}}+\frac{\bm{J}^2}{2V_{11}} \nonumber \\%
&&	+\left[\frac{1}{U_{33}}-\frac{1}{U_{11}}-\frac{4}{V_{11}}\right]\frac{K_3^2}{2}. %
\end{eqnarray}%
The corresponding energy (mass) eigenvalues are
\begin{eqnarray}%
	E=M\Eqa{+}\frac{i(i+1)}{2U_{11}}+\frac{j(j+1)}{2V_{11}} \nonumber \\%
&&	+\left[\frac{1}{U_{33}}-\frac{1}{U_{11}}-\frac{4}{V_{11}}\right]\frac{\kappa^2}{2} 
\label{spectrae}%
\end{eqnarray}%
where $i(i+1),j(j+1)$ and $\kappa$ are the eigenvalues of the Casimir operators 
(\ref{casie}), and the operator $K_3$, respectively. 
The parity is defined by the eigenstate of the following operator
\begin{eqnarray}%
P=e^{i\pi K_3}, 
\end{eqnarray}%
which means that the parity is $(+)$ for even $\kappa$ and $(-)$ for odd $\kappa$. %

\begin{table*}[p]%
\caption{\label{spectrat}The dimensionless quantized energy spectra up to $i,j\le 3$ for 
$\kappa=0$, 
and also some excited states up to $\kappa\le 2$. In the particle classification, 
we give the baryonic descriptions if available, otherwise we only show its spectroscopic 
classification $^{2s+1}{L}_j$.}%
\begin{center}%
\begin{ruledtabular}%
\begin{tabular}{ccccccccc} %
Classification		  & $i$ & $j$ & $\kappa$ &Parity&$\alpha =0$&$\alpha =0.040$&$\alpha 
=0.080$&$\alpha =0.126$\\ \hline \hline%
classical skyrmion	  &   &   &          &      &  69.7195  & 63.6722      & 56.9827       
&  44.4862  \\ \hline%
Deuteron ($^3$S$_1$)	  & 0 & 1 & 0        & $+$  &  69.7227  & 63.6758      & 56.9866     
  &  44.4912  \\ \hline%
NN ($^1$S$_0$)		  & 1 & 0 & 0        & $+$  &  69.7241  & 63.6774      & 56.9886       &  
44.4939  \\ \hline%
($^3$P$_2$)		        & 1 & 2 & 1        & $-$  &  69.7294  & 63.6828      & 56.9943       
&  44.5006  \\ \hline%
N$\Delta$ ($^5$S$_2$)	  & 1 & 2 & 0        & $+$  &  69.7339  & 63.6881      & 57.0003    
   &  44.5089  \\ \hline%
N$\Delta$ ($^3$S$_2$)	  & 2 & 1 & 0        & $+$  &  69.7368  & 63.6913      & 57.0041    
   &  44.5143  \\ \hline%
($^3$P$_2$)		        & 2 & 2 & 1        & $-$  &  69.7387  & 63.6932      & 57.0059       
&  44.5160  \\ \hline%
$\Delta \Delta$ ($^7$S$_3$)& 0 & 3 & 0       & $+$  &  69.7391  & 63.6935      & 57.0062  
     &  44.5162  \\ \hline%
($^5$P$_3$)               & 1 & 3 & 1        & $-$  &  69.7392  & 63.6935      & 57.0060  
     &  44.5156  \\ \hline%
$\Delta \Delta$ ($^1$S$_0$)& 3 & 0 & 0       & $+$  &  69.7475  & 63.7032      & 57.0176  
     &  44.5324  \\ \hline%
($^5$D$_4$)		        & 2 & 4 & 2        & $+$  &  69.7479  & 63.7024      & 57.0153       
&  44.5262  \\ \hline%
($^5$P$_3$)		        & 2 & 3 & 1        & $-$  &  69.7485  & 63.7039      & 57.0177       
&  44.5310  \\ %
\end{tabular}
\end{ruledtabular}
\end{center}%
\label{spectrat}%
\end{table*}%
Taking into account the Finkelstein-Rubinstein constraints for the axially 
symmetric soliton~\cite{bra}, one finds that some combinations of $(i,j)$ are not acceptable.  
The allowed states are~\cite{bra,FR}
\begin{eqnarray}%
&&\left|ii_30 \right>\left|jj_30\right> , \mbox{provided } i+j \mbox{ is odd, } \nonumber \\%
&&\hspace{20mm}(\mbox{for }\kappa=0)\nonumber \\%
&&\frac{1}{\sqrt{2}}[\left|ii_3\kappa \right>\left|jj_3 -2\kappa \right> %
-(-1)^{i+j}\left|ii_3 -\kappa \right>\left|jj_32\kappa \right>]. \nonumber \\%
&&\hspace{20mm}(\mbox{for } \kappa=1, \cdots , \mbox{min}\{i,[j/2]\}) \label{frce}%
\end{eqnarray}%
Therefore $\kappa >0$ is allowed only if $i \geq 1$ and $j \geq 2$. 

\begin{figure}%
\includegraphics[width=7.5cm,keepaspectratio,clip]{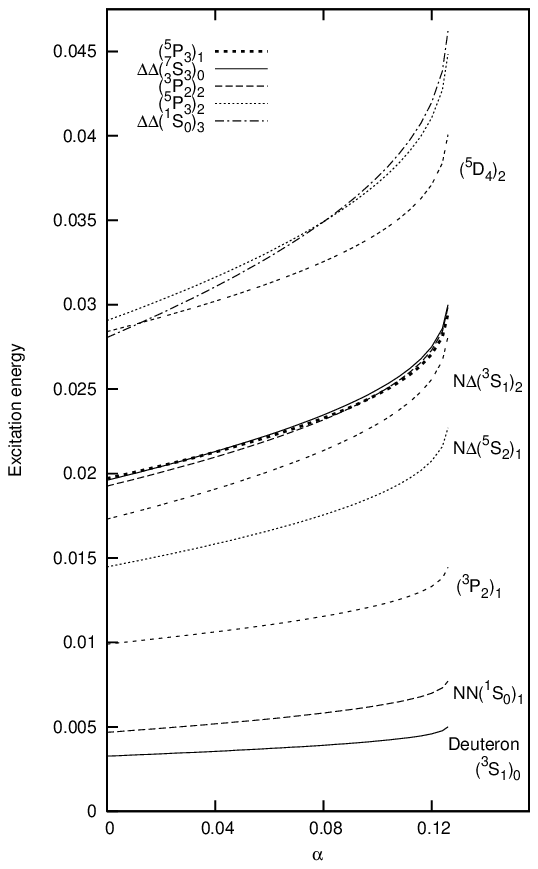}%
\caption{\label{spectdf} The coupling constant dependence \
of the dimensionless mass difference from the classical energy 
are shown. \
For $\alpha \gtrsim 0.127$, there exists no solution. 
We give the baryonic descriptions if available, 
otherwise we only show its spectroscopic classification $(^{2s+1}L_j)_i$, 
whose subscript $i$ means eigenvalue of the isospin operator (\ref{casie}). 
}%
\end{figure}

We show quantized energy spectra for various values of $\alpha$ in Table \ref{spectrat}.  
The mass difference from the classical energy is shown as a function of $\alpha$ in Fig. \ref{spectdf}. %
In Table \ref{spectrat} (and Fig.\ref{spectdf}), we show the results for $i,j\le 3$ with $\kappa=0$, 
and also some excited states for $\kappa\le 2$.%
We present a spectroscopic classification $ ^{2s+1}L_j$ in Table \ref{spectrat}.  
For $\kappa=0$, the system is in {\it S-state}, and then $j$ equals to the intrinsic spin $s$. 
For $\kappa>0$ state, the orbital angular momentum is chosen to be its lowest value 
so that it is consistent with the quantum number $j$, the parity and the fact that $s\le 3$ in the six-quark picture. 
It is seen that the mass difference increases monotonically with increasing $\alpha$. 
Fig.~\ref{pfpcf} (and also Figs.\ref{edenf},\ref{bdenf}) 
implies that the strong gravity makes the size of the skyrmion smaller which 
makes the inertial moment smaller, resulting in increase in the mass difference. 
In the collective quantization, the skyrmion can be quantized as a slowly rotating 
rigid body and the mass difference is interpreted as a consequence of the rotational 
kinetic energy. Thus the gravity works for increasing the kinetic energy of 
the skyrmion. In the naive quark model picture, the mass difference is ascribed to 
the hyperfine splittings. The increase in the mass difference  
may imply that due to the reduction of the distance between quarks,  
the effects of the hyperfine splittings become dominant by the gravity~\cite{glashow}.   

From the data of the Table \ref{spectrat}, one can straightforwardly obtain the mass value
in MeV unit, by multiplying $f_{\pi}/e$ for any parameter choice $(f_\pi,e)$.
With $f_\pi=108$ MeV, $e=4.84$, which is used in Ref.~\cite{bra}, our obtained mass value 
in the case of $\alpha=0$ is approximately $6 \% $ lower than the value obtained in Ref.~\cite{bra}.
There are two possible reasons for that.
Firstly, the Lagrangian in Ref.~\cite{bra} includes the mass term and hence it slightly 
enhances the mass value.
Thus we also performed the same analysis including the mass term, which reduced the error from 
$6\%$ to $0.5\%$ in the mass. Secondly, the coordinate system they used is cylindrical while we 
adopted the spherical coordinates. 
Since the cylindrical coordinate requires much larger number of grid point to obtain torus-shape 
solutions than the spherical, we suspect that their results are not fully convergent, 
producing a slightly larger mass. 

\section{Electromagnetic properties}

In this section we shall investigate the gravity effects to the various observables, {\it i.e.}
the mean charge radius $\left< r^2 \right>_{d}^{1/2}$, magnetic moment $\mu_d$, 
the quadrupole moment $Q$, and the transition moment $\mu_{d \rightarrow np}$, 
which is associated with the process $\gamma d \rightarrow {^1S_{0}}$. %

Let us introduce the electromagnetic current %
in the Skyrme model $J^{({\rm em})}_{\mu}(x)$ which consists of the isoscalar and isovector part, %
\begin{eqnarray}%
J^{({\rm em})}_{\mu}(x)=\frac{1}{2}B_{\mu}(x)+I_{\mu}^3(x)\,, \label{elemagce}%
\end{eqnarray}%
where $B_{\mu}(x)$ is baryon current density given in Eq.(\ref{bcde}) and %
$I_{\mu}^3(x)$ is the third component of the isospin current density, defined by %
\begin{eqnarray}%
I_{\mu}^a (x)\Eqa{=}-e f_{\pi}^3\frac{i}{8}\bigl( %
	g^{\mu \nu}{\rm Tr} [U^{-1} [{\textstyle \frac{\bm{\tau}_i}{2}},U ] U^{-1}\partial_{\mu}U] \nonumber\\ %
&&	+g^{\mu \rho}g^{\nu \sigma} \ %
	{\rm Tr} [ U^{-1}[{\textstyle \frac{\bm{\tau}_i}{2}},U ] U^{-1}\partial_{\nu}U ] \nonumber \\ %
&&		 \times [U^{-1}\partial_{\mu}U,U^{-1}\partial_{\nu}U] \bigr) .%
\end{eqnarray}%
Inserting the dynamical field $\hat{U}$ in Eq.(\ref{rote}) into Eq.(\ref{elemagce}) %
gives electromagnetic current operator $\hat{J}_{\mu}^{({\rm em})}$. %

To estimate the expectation value of these quantum operators, we describe the 
quantum spin states (\ref{frce}) in terms of the products of the rotation matrices
\begin{eqnarray}%
\left< A | i i_3 k_3 \right> \Eqa{=}\left(\frac{2i+1}{8\pi^2}
\right)^{\frac{1}{2}}D^i(i\tau^2 A^{\dagger})_{k_3 i_3}, \label{wdfAe}\\%
\left< B | j j_3 l_3 \right> \Eqa{=}\left(\frac{2j+1}{8\pi^2}
\right)^{\frac{1}{2}}D^j(i\tau^2 B^{\dagger})_{j_3 l_3}. \label{wdfBe}%
\end{eqnarray}%
where $D^i(A)_{m,m'}$ is well known Wigner $D$ function.
The computation can be easily performed by the following integration formula~\cite{qtoam}%
\begin{eqnarray}%
\int dB && \hspace{-3mm} D^{i}(B)^{*}_{m_1m_1'}D^{j}(B)_{m_2m_2'}D^{k}(B)_{m_3m_3'} \nonumber \\ %
\Eqa{=}\frac{8 \pi^2}{2k+1}C_{im_1jm_2}^{km_3}C_{im_1'jm_2'}^{km_3'}, \label{wdie}%
\end{eqnarray}%
where $C_{im_1jm_2}^{km_3}$ is a Clebsch-Gordan coefficient. %

\begin{figure}%
\includegraphics[width=7.5cm,keepaspectratio,clip]{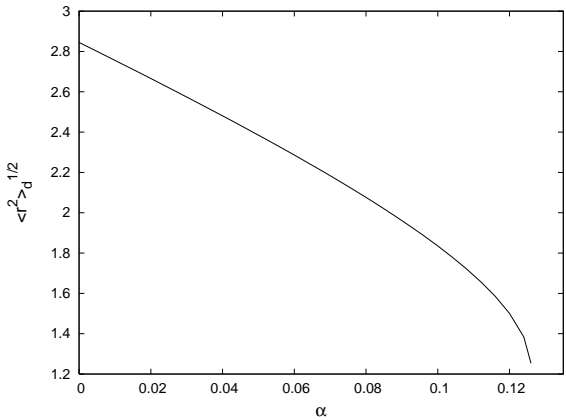}%
\caption{\label{rmsrf} The coupling constant dependence of the \
dimensionless mean charge radius (\ref{rmsre}). %
}%
\end{figure}

The deuteron charge radius $\left<r^2\right>^{1/2}_d$ is defined as the square root of %
\begin{eqnarray}%
\left<r^2\right>_d \equiv \bigl<d \bigl|\int d^3r r^2 \hat{J}_0^{({\rm em})}(\bm{r},t)\bigr|d\bigr> ,%
\end{eqnarray}%
where $\left.\left.\right|d\right>$ represents spin state of the deuteron. %
In this case,  as only the isoscalar part of $\hat{J}_{0}^{({\rm em})}$ contributes to the matrix 
element, the integration is straightforward.
\begin{eqnarray}%
\left<r^2\right>_d \Eqa{=} \frac{\pi}{(e f_{\pi})^2} \int dx d\theta \sqrt{-g} x^2 B^0(r,\theta) \nonumber \\ %
\Eqa{=} -\frac{1}{\pi (e f_{\pi})^2}\int dx d\theta x^2 \sin^2{F}\sin{\Theta} \nonumber \\ %
	&& \hspace{20mm} \times (\partial_{x} F \partial_{\theta}\Theta-\partial_{\theta} F \partial_{x}\Theta  ) .\label{rmsre}%
\end{eqnarray}%
Fig.\ref{rmsrf} shows the $\alpha$ dependence of the mean charge radius. %
Due to the attractive effect of the gravity, it decrease with increasing $\alpha$. %

\begin{figure}%
\includegraphics[width=7.5cm,keepaspectratio,clip]{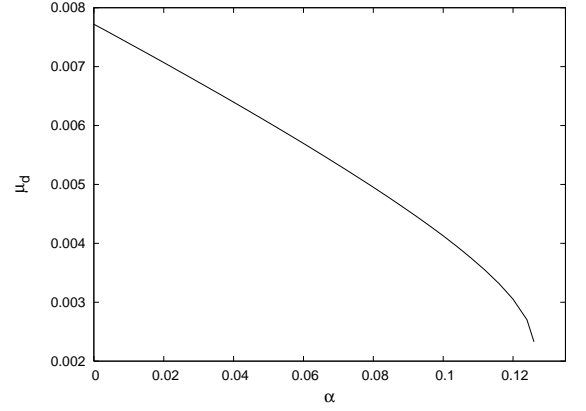}%
\caption{\label{mudf} The coupling constant dependence of the dimensionless magnetic moment (\ref{mude}).%
}%
\end{figure}

\begin{figure}%
\includegraphics[width=7.5cm,keepaspectratio,clip]{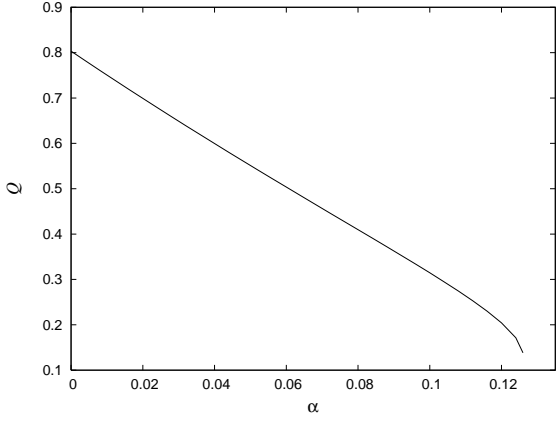}%
\caption{\label{qdplmf} The coupling constant dependence of the \
dimensionless quadrupole moment (\ref{qdplme}).%
}%
\end{figure}

The isoscalar part of the magnetic moments is expressed in term of electromagnetic current as %
\begin{eqnarray}%
\hat{\mu}_i=\frac{1}{2}\int d^3r\sqrt{-g} \epsilon_{ijk}r_j\hat{J}_0^{{\rm (em)}}(\bm{r},t). \label{magme}%
\end{eqnarray}%
Inserting the dynamical field $\hat{U}$ in Eq.(\ref{rote}) into Eq.(\ref{magme}), one finds the magnetic momentum operator as %
\begin{eqnarray}%
\hat{\mu}_i \bigr|_{I=0} = \frac{1}{2} \{ R_{ij}(B)^T, \mu_{jk}a_k+\mu'_{jk}b_k \}, %
\end{eqnarray}%
where %
\begin{eqnarray}%
\mu_{jk} \Eqa{=} \frac{1}{(ef_{\pi})^2}\int d^3 x\frac{1}{2} \varepsilon_{jlm}x_l %
\frac{1}{2}{\rm Tr}[U^{\dagger}[\frac{1}{2}\bm{\tau}_k,U]C_m] , \\ %
\mu_{jk}'\Eqa{=} \frac{1}{(ef_{\pi})^2}\int d^3 x\frac{1}{2} \varepsilon_{jlm}x_l %
\frac{1}{2}{\rm Tr}[U^{\dagger}(-i\bm{x}\times \bm{\nabla})_k U \nonumber \\ %
&& \hspace{50mm} \times C_m(\bm{x})], %
\end{eqnarray}%
and $C_m$ is %
\begin{eqnarray}%
C_m(\bm{x})=\frac{i}{8\pi ^2}\varepsilon_{mnp}(U^{\dagger}\partial_n 
U)(U^{\dagger}\partial_p U). %
\end{eqnarray}%
The $R_{ij}$ is the rotation matrix (\ref{rotme}) and $a_i$ and $b_i$ are defined as Eq.(\ref{tope}). %
The anti-commutator relation in Eq.(\ref{magme}) guarantees $\hat{\mu}_{i}$ to be Hermitian operator. %
$\mu_{ij}$ and $\mu_{jk}'$ are diagonal and furthermore they satisfy %
$\mu_{11}=\mu_{22}=0$ , $\mu_{11}'=\mu_{22}'$ and $\mu_{33}'=-2\mu_{33}$. %
With these relations and using the body-fixed operator (\ref{bfoe}) , the coordinate-fixed operator (\ref{cfoe}) , %
 the relation of the moment of inertia components (\ref{midie}) , and %
the constraint of Eq.(\ref{conste}) in Eq.(\ref{magme}), one can get %
\begin{eqnarray}%
\hat{\mu}_i\bigr|_{I=0}=-\frac{\mu_{11}'}{V_{11}}\bm{J}_l+\mbox{terms proportional to }K_3.  %
\end{eqnarray}%
Therefore we only use the component of $\mu_{11}'$, %
\begin{eqnarray}%
\mu_{11}'=-\frac{\pi}{4(e f_{\pi})^2}\int dxd\theta \sqrt{-g} x^2(\cos^2{\theta}+1)B_0(x,\theta).%
\end{eqnarray}%
And the $V_{11}$ is moment of inertia given in Eq.(\ref{v11e}). %
The moments $\mu_d$ of deuteron is defined as the expectation value of the $\hat{\mu}_3$ with $j_3=1$%
\begin{eqnarray}%
\mu_d=\bigl<\hat{\mu}_3\bigr>=-\frac{\mu_{11}'}{V_{11}}.\label{mude}%
\end{eqnarray}%
In Fig.\ref{mudf} $\alpha$ dependence of the $\mu_d$ is shown. %
It decreases with increasing $\alpha$, but the effect of gravity is more evident 
compared with that of $B=1$~\cite{np2}.  
In the non-relativistic nuclear physics point of view, the magnetic moment of the 
deuteron consists of the sum of the magnetic moment of the proton plus neutron and some 
correction terms related to the {\it D-state probability} of the deuteron. 
This result indicates that the assumption that the deuteron is almost S-state 
(contributions to the other states are only $\sim 5\%$) may no longer valid 
under the strong gravitational field. 
Therefore, we speculate that the effect of gravity is
apparent on the change of such non-central components of the nuclear force. 
This will be more evident by examining the quadrupole moment.
The quadrupole moment is given by
\begin{eqnarray}%
\hat{Q}_{ij} \Eqa{=} \int 
d^3r\sqrt{-g}(3r_ir_j-r^2\delta_{ij})\hat{J}_0^{(em)}(\bm{r},\theta), \label{qelme}%
\end{eqnarray}%
which is the same as the magnetic moment. Inserting (\ref{rote}) into Eq.(\ref{qelme}) one obtains 
the quadrupole momentum operator %
\begin{eqnarray}%
\hat{Q}_{ij} \bigr|_{I=0} \Eqa{=} R_{ia}(B)^T Q_{ab}R_{bj}(B), %
\end{eqnarray}%
where
\begin{eqnarray}%
Q_{ab}=\frac{1}{2(ef_{\pi})^2}\int d^3x\sqrt{-g}(3x_a x_b-x^2\delta _{ab})B_0(x,\theta). 
\label{qabe}%
\end{eqnarray}%
$Q_{ij}$ satisfy $Q_{11}=Q_{22}$. %
And the symmetry relation for the quadrupole moments reduces the expression to %
\begin{eqnarray}%
\hat{Q}_{ij}\bigr|_{I=0}=Q_{33}[\frac{3}{2}R_{i3}(B)^{T}R_{3j}(B)-\frac{1}{2}\delta_{ij}]. \label{q33e}%
\end{eqnarray}%
Thus we only need the component of $Q_{33}$ %
\begin{eqnarray}%
Q_{33} \Eqa{=}\frac{\pi}{(e f_{\pi})^2}\int dx d \theta\sqrt{-g} x^2(3\cos^2{\theta}-1)B^0(x,\theta). %
\end{eqnarray}%
The deuteron quadrupole moment is defined as the expectation value of $\hat{Q}_{33}$ with $j_3=1$ %
\begin{eqnarray}%
Q\Eqa{=}\bigl< \hat{Q}_{33} \bigr>, \nonumber \\%
\Eqa{=} Q_{33} \biggl[\frac{3}{2}\bigl<d,j_3'=1 \bigr| R_{33}^T(B)R_{33}(B) 
\bigl|d,j_3=1\bigr>-\frac{1}{2} \biggr].\nonumber \\ %
&& \label{qdplmmate} %
\end{eqnarray}%
Now let us evaluate the matrix element. %
The rotating matrix $R_{33}$ is represented as $R_{33}(B)=D^1(B)_{00}$ with the Wigner $D$ function. %
The product of two $D$-functions can be expanded in the following series \cite{qtoam}%
\begin{eqnarray}%
&& \hspace{-4mm}D^{J_1}(B)_{M_1N_1}D^{J_2}(B)_{M_2N_2} \nonumber \\ %
\Eqa{=}\sum^{J_1+J_2}_{J=|J_1-J_2|} 
\sum_{MN}C^{JM}_{J_1M_1J_2M_2}D^J(B)_{MN}C^{JM}_{J_1N_1J_2N_2}. \label{wdpe}%
\end{eqnarray}%
The matrix element of $D^1(B)_{00}$ derived from Eqs.(\ref{wdfBe}) and (\ref{wdie}) %
and the relation of the $D^l(i\tau_2)_{0m}=(-1)^{l}\delta_{m,0}$, %
\begin{eqnarray}%
\bigl<j'j_3'l_3' \bigr| D^1(B)_{00} \bigl|jj_3l_3 \bigr> %
= -\bigl(\frac{2j'+1}{2j+1} \bigr)^{\frac{1}{2}}C_{j'l'_310}^{jl_3}C_{j'j'_310}^{jj_3}. \label{mer33e}%
\end{eqnarray}%
The deuteron state is the one with $j_3=1$. Therefore inserting $i=0$, $j=1$, $\kappa=0$ 
and using the constraint in Eq.(\ref{conste}), the matrix element in Eq(\ref{qdplmmate}) is evaluated as %
\begin{eqnarray}%
Q=-\frac{1}{5}Q_{33}. \label{qdplme}%
\end{eqnarray}%
In Fig.\ref{qdplmf} $\alpha$ dependence of the $Q$ is shown. %
As in the case of the magnetic moment, the quadrupole moment 
significantly decreases with increasing $\alpha$, which 
suggests that the D-state probability is strongly affected by the gravity. 

\begin{figure}%
\includegraphics[width=7.5cm,keepaspectratio,clip]{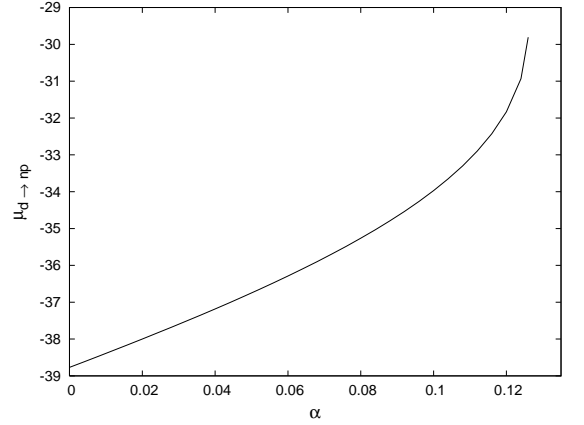}%
\caption{\label{mu_dnpf} The coupling constant dependence of the \
dimensionless transition moment (\ref{mudnpe}). %
}%
\end{figure}

The transition moment $\mu_{d\rightarrow np}$ for photodisintegration of the deuteron 
into the isovector $^1S_0$ state is defined by the magnetic moment operator $\hat{\mu}_3$. %
It is evaluated in terms of matrix element between the $j_3=0$ state and $i_3=0$ state. %
\begin{eqnarray}%
\mu_{d\rightarrow np}=\bigl{<} {^1S_0} , i_3=0\bigr|\hat{\mu}_3\bigl|d,j_3=0\bigr>.\label{trae} %
\end{eqnarray}%
Inserting the dynamical field in Eq.(\ref{rote}) into the isovector part of $\hat{\mu}_{i}$, one can obtain %
\begin{eqnarray}%
\hat{\mu}_{i} \bigr|_{I=1}=-\frac{1}{2}R_{3j}(A)W_{jk}R_{ki}(B) %
\end{eqnarray}%
where $W_{jk}$ is the moment of inertia tensor defined in Eq.(\ref{moiwe}). %
From Eq.(\ref{midie}), one can see that $W_{33}$ is the only non-zero component and $W_{33}=2U_{33}$, and hence %
\begin{eqnarray}%
\hat{\mu}_{3}\big|_{I=0}=-U_{33}R_{33}(A)R_{33}(B).%
\end{eqnarray}%
Evaluating the matrix elements of (\ref{trae}) using (\ref{mer33e}) and the element of the %
isospace which is derived in the similar way in Eq.(\ref{mer33e}), one can get %
\begin{eqnarray}%
\mu_{d\rightarrow np}=-\frac{1}{3}U_{33}. \label{mudnpe}%
\end{eqnarray}%
In Fig.\ref{mu_dnpf} the coupling constant dependence of $\mu_{d\rightarrow np}$ is shown. %
As expected, it also decreases with increasing $\alpha$. 
Since the strong gravity reduces the transition moment significantly, it may be possible 
to determine the gravitational constant by observing the variation in $\mu_{d\rightarrow np}$. 
It is interesting that the decay rates are reduced by the gravitational 
effects whether the interaction is strong or electromagnetic, which means 
the gravity works as a stabilizer of baryons. 
\section{conclusions}

In this article we have investigated axially symmetric $B=2$ skyrmions coupled to gravity. 
Performing collective quantization, we obtained the static observables of the dibaryons. 
The rotational and isorotational modes were quantized in the same manner as the skyrmion 
without gravity. It was shown how the static properties of dibaryons such as masses, 
charge densities, magnetic moments were modified by the gravitational interaction. 

The dependence of the energy density and mean radius on the coupling constant showed that 
the soliton shrinks as $\alpha$ become larger which reflects the attractive feature of 
gravity. The mass difference between dibaryons also becomes larger as increasing $\alpha$. 
These observations can be interpreted that shrinking the skyrmion reduces  
the inertial momenta and hence induces the large mass difference between dibaryon spectra.
 
In the collective quantization, the skyrmion can be quantized as a slowly rotating 
rigid body and the various dibaryon spectra are regarded as the rotational bands of the 
classical skyrmion. Thus the large mass difference means that the gravity works 
for increasing the rotational kinetic energy of the skyrmion. 
The magnetic moment is reduced significantly as increasing $\alpha$ compared to  
that of $B=1$. We also calculated the quadrupole moment. For the transition moments 
of the deuteron, we found that the gravity works as a stabilizer. 
Thus, it may be possible to determine the gravitational constant by the measurement of 
the various decay rates of the dibaryonic objects. 

Throughout the paper, we consider $\alpha$ as a free parameter and studied skyrmion 
spectra in the strong coupling limit. 
We found that the effects of gravity on the observables estimated here are manifested 
only in such a large coupling constant. 
Some theories such as scalar-tensor gravity theory~\cite{brans}
and theories with extra dimensions discuss the time variation of 
the gravitational constant~\cite{marciano}. There may have been an 
epoch in the early universe where the gravitational effects on nucleons 
were significant. %

Let us note that in Ref.\cite{leese} the authors computed the quantum 
correction to the $B=2$ skyrmion by considering the moduli space
of instanton-generated two-skyrmions in the attractive channel 
which has a larger dimension (${\cal M}_{10}$) than the moduli space of 
the deuteron (${\cal M}_{8}$), which greatly improved the deuteron observables. 
Thus, going beyond the collective coordinate approximation provides a much greater 
correction to the observables than gravitational effects. 

As a future work, the analysis of the electromagnetic form factors of the 
deuteron is now under consideration. 
The SU(3) extension and the analysis of dihyperon coupled to  
gravity will be also interesting. 

\section{Acknowledgement}
We would like to thank Rajat K. Bhaduri for drawing our attention to this subject and 
useful comments. 

\appendix %

\section{Field eqations}\label{feqsa}%
For the numerical calculation, we employ the radial coordinate %
\begin{eqnarray}%
	\chi = \frac{x}{1+x} %
\end{eqnarray}%
to map infinity to $\chi=1$.  %
The regularity requires the condition %
\begin{eqnarray}%
m(\chi,0)=l(\chi,0).%
\end{eqnarray} %
Following the work of Ref.~\cite{ioa} we introduce the metric function $g$ as %
\begin{eqnarray}%
g(\chi,\theta)=\frac{m(\chi,\theta)}{l(\chi,\theta)} . %
\end{eqnarray}%
Then the boundary conditions for $g$ are given by %
\begin{eqnarray}%
g(0,\theta)\Eqa{=}1 \ , \ g(\infty,\theta)=1 , \\%
g(\chi,0)\Eqa{=}1 \ , \ g(\chi,\frac{\pi}{2})=0 . %
\end{eqnarray}%
The filed equations for the profile function $F(\chi,\theta)$ and $\Theta(\chi,\theta)$ 
are derived as  %
\begin{eqnarray}%
&&\hspace{-4mm} \left(\frac{\delta \sqrt{-g}\mathcal{L}_S}{\delta 
F}\right)\frac{1}{f_{\pi}^2}\frac{4}{\sqrt{l}\sin{\theta}} \nonumber \\%
\Eqa{=} \chi^2(1-\chi)^2(F_{,\chi \chi}+{\textstyle 
\frac{l_{,\chi}}{2l}}F_{,\chi})+2\chi(1-\chi)^2F_{,\chi} \nonumber \\ %
&&+F_{,\theta \theta}+F_{,\theta}({\textstyle \frac{l_{,\theta}}{2l}}+\cot{\theta}) 
\nonumber \\%
&&  -\{ \chi^2(1-\chi)^2 \Theta _{,\chi}^2 +\Theta _{,\theta}^2 \}\sin{F} \cos{F} 
\nonumber \\%
&&-{\textstyle \frac{n^2 m}{l \sin^2{\theta}}}\sin{F} \cos{F} \sin^2{\Theta} \nonumber \\%
&& +4{\textstyle \frac{f}{m}}\sin^2{F}\Bigl[(1-\chi)^4  %
  \bigl\{ ({\textstyle \frac{f_{,\chi}}{f}}+{\textstyle \frac{l_{,\chi}}{2l}} %
  -{\textstyle \frac{m_{,\chi}}{m}})\Theta_{,\theta} \nonumber \\ %
&&\hspace{4mm} -({\textstyle \frac{f_{,\theta}}{f}} %
  +{\textstyle \frac{l_{,\theta}}{2l}}-{\textstyle 
\frac{m_{,\theta}}{m}}+\cot{\theta})\Theta_{,\chi} \nonumber \\%
&&\hspace{4mm} +\cot{F}(F_{,\chi}\Theta_{,\theta}-F_{,\theta}\Theta_{,\chi})\bigr\} %
  (F_{,\chi}\Theta_{,\theta}-F_{,\theta}\Theta_{,\chi}) \nonumber \\%
&&\hspace{4mm} +\bigl\{ \Theta_{,\theta}(F_{,\chi 
\chi}\Theta_{,\theta}+F_{,\chi}\Theta_{,\chi \theta} %
  -F_{,\chi \theta}\Theta_{,\chi}-F_{,\theta}\Theta_{,\chi \chi})\nonumber \\%
&&\hspace{4mm} -\Theta_{,\chi}(F_{,\chi \theta}\Theta_{,\theta}+F_{,\chi}\Theta_{,\theta 
\theta} %
  -F_{,\theta \theta}\Theta_{,\chi}-F_{,\theta}\Theta_{,\chi \theta}) \bigr\}\nonumber \\%
&&\hspace{4mm} \times (1-\chi)^4 -2(1-\chi)^3 
\Theta_{,\theta}(F_{,\chi}\Theta_{,\theta}-F_{,\theta}\Theta_{,\chi}) \Bigr]\nonumber \\%
&&+{\textstyle \frac{4n^2f}{l \sin^2{\theta}}}\sin^2{F}\sin^2{\Theta}\Bigl[(1-\chi)^4 %
  \bigl\{ F_{,\chi \chi} \nonumber \\%
&&\hspace{4mm} +F_{,\chi}^2\cot{F}+2F_{,\chi}\Theta_{,\chi}\cot{\Theta} %
  +({\textstyle \frac{f_{,\chi}}{f}}-{\textstyle \frac{l_{,\chi}}{2l}})F_{,\chi} \nonumber
 \\%
&&\hspace{4mm} -2\sin{F}\cos{F}\Theta_{,\chi}^2 \bigr\}-2(1-\chi)^3F_{,\chi} \nonumber \\%
&&\hspace{4mm} +{\textstyle \frac{(1-\chi)^2}{\chi^2}} \bigl\{F_{,\theta \theta} %
+F_{,\theta}^2\cot{F}+2F_{,\theta}\Theta_{,\theta}\cot{\Theta} \nonumber \\%
&&\hspace{4mm} +({\textstyle \frac{f_{,\theta}}{f}}-{\textstyle 
\frac{l_{,\theta}}{2l}}-\cot{\theta})F_{,\theta}
 -2\Theta_{\theta}^2\sin{F}\cos{F} \bigr\} \Bigr], \label{feqfe}%
\end{eqnarray}%
\begin{eqnarray}%
&&\hspace{-4mm} \left(\frac{\delta \sqrt{-g}\mathcal{L}_S}{\delta 
\Theta}\right)\frac{1}{f_{\pi}^2}\frac{4}{\sqrt{l}\sin{\theta}}\frac{1}{\sin^2{F}} 
\nonumber \\%
\Eqa{=}\chi^2(1-\chi)^2(\Theta_{,\chi \chi}+{\textstyle 
\frac{l_{,\chi}}{2l}}\Theta_{,\chi})+2\chi(1-\chi)^2\Theta_{,\chi} \nonumber \\%
&&+\Theta_{,\theta \theta}+\Theta_{,\theta}({\textstyle 
\frac{l_{,\theta}}{2l}}+\cot{\theta}) \nonumber \\%
&&+2\{ \chi^2(1-\chi)^2F_{,\chi}\Theta_{,\chi}+F_{,\theta}\Theta_{,\theta} \} \cot{F} 
\nonumber \\%
&&-{\textstyle \frac{n^2 m}{l \sin^2{\theta}}}\sin{\Theta} \cos{\Theta} \nonumber \\ %
&&+4 {\textstyle \frac{f}{m}}\Bigl[(1-\chi)^4 \bigr\{ %
  ({\textstyle \frac{f_{,\theta}}{f}}+{\textstyle \frac{l_{,\theta}}{2l}} %
  -{\textstyle \frac{m_{,\theta}}{m}}+ \cot{\theta})F_{,\chi} \nonumber \\%
&&\hspace{4mm} -({\textstyle \frac{f_{,\chi}}{f}}+{\textstyle \frac{l_{,\chi}}{2l}} %
  -{\textstyle \frac{m_{,\chi}}{m}})F_{,\theta} \bigr\} 
(F_{,\chi}\Theta_{,\theta}-F_{,\theta}\Theta_{,\chi}) \nonumber \\%
&&\hspace{4mm} + \{ F_{,\chi}(F_{,\chi \theta}\Theta_{,\theta}+F_{,\chi}\Theta_{,\theta 
\theta} %
  -F_{,\theta \theta}\Theta_{,\chi}-F_{,\theta}\Theta_{,\chi \theta}) \nonumber \\%
&&\hspace{4mm} -F_{,\theta}(F_{,\chi \chi}\Theta_{,\theta}+F_{,\chi}\Theta_{,\chi \theta} 
  -F_{,\chi \theta}\Theta_{,\chi}-F_{,\theta}\Theta_{,\chi \chi}) \}\nonumber \\%
&&\hspace{4mm} \times (1-\chi)^4 +2(1-\chi)^3 
F_{,\theta}(F_{,\chi}\Theta_{,\theta}-F_{,\theta}\Theta_{,\chi})\Bigr] \nonumber \\%
&&+{\textstyle \frac{4n^2f}{l \sin^2{\theta}}}\sin^2{F}\sin^2{\Theta} %
  \Bigl[(1-\chi)^4 \{ \Theta_{,\chi \chi} \nonumber \\%
&&\hspace{4mm} +\Theta_{,\chi}^2\cot{\Theta}+4F_{,\chi}\Theta_{,\chi}\cot{F} \nonumber \\%
&&\hspace{4mm} +({\textstyle \frac{f_{,\chi}}{f}}-{\textstyle 
\frac{l_{,\chi}}{2l}})\Theta_{,\chi} \bigr\} %
  -2(1-\chi)^3\Theta_{,\chi} \nonumber \\%
&&\hspace{4mm} +{\textstyle {\textstyle \frac{(1-\chi)^2}{\chi^2}}}\{\Theta_{,\theta 
\theta} %
  +\Theta_{,\theta}^2\cot{\Theta} %
  +4F_{,\theta}\Theta_{,\theta}\cot{F} \nonumber \\%
&&\hspace{4mm} +({\textstyle \frac{f_{,\theta}}{f}}-{\textstyle \frac{l_{,\theta}}{2l}} %
  -\cot{\theta})\Theta_{,\theta}\bigr\} \Bigr] \nonumber \\%
&&{\textstyle -\frac{4n^2f}{l \sin^2{\theta}}} \sin{\Theta} \cos{\Theta} %
  \left[(1-\chi)^4 F_{,\chi}^2+{\textstyle \frac{(1-\chi)^2}{\chi^2}}F_{,\theta}^2 
\right].\label{feqce} %
\end{eqnarray}%
From Einstein equations, the field equations for the metric functions are derived.  %
We diagonalized those equations to the 2nd-derivative of $\chi$ and $\theta$ for each 
metric function %
\begin{eqnarray}%
&&\hspace{-4mm} {\textstyle \frac{m}{f^2}}x^2G_{00}+x^2G_{11}+G_{22}+{\textstyle 
\frac{m}{l \sin^2{\theta}}}G_{33} \nonumber \\%
&&\hspace{-4mm}-2 \alpha \Bigl[{\textstyle 
\frac{m}{f^2}}\hat{r}^2T_{00}+\hat{r}^2T_{11}+T_{22} %
  +{\textstyle \frac{m}{l\sin^2{\theta}}}T_{33}\Bigr] \nonumber \\%
\Eqa{=} \chi^2(1-\chi)^2{\textstyle \frac{f_{,\chi 
\chi}}{f}}-\chi^2(1-\chi)^2\bigl({\textstyle \frac{f_{,\chi}}{f}}\bigr)^2 %
  +2\chi(1-\chi)^2{\textstyle \frac{f_{,\chi}}{f}} \nonumber \\%
&&+{\textstyle \frac{f_{,\theta  \theta}}{f}}-\bigl({\textstyle 
\frac{f_{,\theta}}{f}}\bigr)^2 %
  +\cot{\theta}{\textstyle \frac{f_{,\theta}}{f}} \nonumber \\%
&&+{\textstyle \frac{1}{2}}\chi^2(1-\chi)^2{\textstyle \frac{f_{,\chi}}{f}}{\textstyle 
\frac{l_{,\chi}}{l}} %
  +{\textstyle \frac{1}{2}}{\textstyle \frac{f_{,\theta}}{f}}{\textstyle 
\frac{l_{,\theta}}{l}} \nonumber \\%
&&-\alpha\bigl[2{\textstyle 
\frac{f}{m}}(1-\chi)^4(F_{,\chi}\Theta_{,\theta}-F_{,\theta}\Theta_{,\chi})^2 \sin^2{F} 
\nonumber \\%
&&\hspace{4mm} +{\textstyle \frac{2n^2f}{l\sin^2{\theta}}} \{ 
(1-\chi)^4(F_{,\chi}^2+\Theta_{,\chi}^2 \sin^2{F}) \nonumber \\%
&&\hspace{4mm} +{\textstyle \frac{(1-\chi)^2}{4 \chi^2}}(F_{,\theta}^2+\Theta_{,\theta }^2
 \sin^2{F}) \} %
  \sin^2{F} \sin^2{\Theta}\bigr] ,\label{feqmme} %
\\
&&\hspace{-4mm} {\textstyle \frac{m}{l \sin^2{\theta}}}G_{33} %
-2 \alpha \Bigl[{\textstyle \frac{m}{l\sin^2{\theta}}}T_{33}\Bigr] \nonumber \\%
\Eqa{=} {\textstyle \frac{1}{2}}\chi^2(1-\chi)^2{\textstyle \frac{m_{,\chi \chi}}{m}} %
  +{\textstyle \frac{1}{2}}\chi(1-\chi)(1-2\chi){\textstyle \frac{m_{,\chi}}{m}} \nonumber
 \\ %
&&-{\textstyle \frac{1}{2}}\chi^2(1-\chi)^2\bigl({\textstyle \frac{m_{,\chi}}{m}}\bigr)^2 
  +{\textstyle \frac{1}{2}}{\textstyle \frac{m_{,\theta  \theta}}{m}} %
  -{\textstyle \frac{1}{2}}\bigl({\textstyle \frac{m_{,\theta}}{m}}\bigr)^2 \nonumber \\%
&&+{\textstyle \frac{1}{4}}\chi^2(1-\chi)^2\bigl({\textstyle \frac{f_{,\chi}}{f}}\bigr)^2
  +{\textstyle \frac{1}{4}}\bigl({\textstyle \frac{f_{,\theta}}{f}}\bigr)^2 \nonumber \\%
&&+\frac{\alpha}{4} \bigl[\chi^2(1-\chi)^2(F_{,\chi}^2+\Theta_{,\chi}^2 \sin^2{F}) 
\nonumber \\ %
&&\hspace{4mm} +F_{,\theta}^2+\Theta_{,\theta }^2 \sin^2{F} %
  -{\textstyle \frac{n^2m}{l\sin^2{\theta}}}\sin^2{F}\sin^2{\Theta} \nonumber \\ %
&&\hspace{4mm} +4{\textstyle \frac{f}{m}} %
  (1-\chi)^4(F_{,\chi}\Theta_{,\theta}-F_{,\theta}\Theta_{,\chi})^2 \sin^2{F} \nonumber 
\\%
&&\hspace{4mm} -{\textstyle \frac{4n^2f}{l\sin^2{\theta}}}\sin^2{F} \sin^2{\Theta} %
  \bigl\{ (1-\chi)^4(F_{,\chi}^2+\Theta_{,\chi}^2 \sin^2{F}) \nonumber \\%
&&\hspace{4mm} 
+{\textstyle \frac{(1-\chi)^2}{\chi^2}}(F_{,\theta}^2+\Theta_{,\theta }^2 \sin^2{F}) 
\bigr\} \bigr] \,,\label{feqmle}%
\\
&&\hspace{-4mm} x^2G_{11}+G_{22}-2\alpha \Bigl[x^2T_{11}+T_{22}\Bigr] \nonumber \\%
\Eqa{=} {\textstyle \frac{1}{2}}\chi^2(1-\chi)^2{\textstyle \frac{l_{,\chi \chi}}{l}} %
  -{\textstyle \frac{1}{4}}\chi^2(1-\chi)^2({\textstyle \frac{l_{,\chi}}{l}})^2 \nonumber 
\\%
&&+\chi(1-\chi)({\textstyle \frac{3}{2}}-\chi){\textstyle \frac{l_{,\chi}}{l}} %
  +{\textstyle \frac{1}{2}}{\textstyle \frac{l_{,\theta  \theta}}{l}} %
  -{\textstyle \frac{1}{4}}({\textstyle \frac{l_{,\theta}}{l}})^2+\cot{\theta}{\textstyle 
\frac{l_{,\theta}}{l}} \nonumber \\%
&&-\alpha \bigl[2{\textstyle 
\frac{f}{m}}(1-\chi)^4(F_{,\chi}\Theta_{,\theta}-F_{,\theta}\Theta_{,\chi})^2 
\sin^2{F}\nonumber \\%
&&\hspace{4mm} -{\textstyle 
\frac{n^2m}{2l\sin^2{\theta}}}\sin^2{F}\sin^2{\Theta}\bigr]\,.\label{feqmfe} %
\end{eqnarray}%
%

\end{document}